\documentclass[final,5p,twocolumn]{elsarticle}

\usepackage{amsmath,amssymb,mathtools}
\usepackage[colorinlistoftodos]{todonotes}
\presetkeys{todonotes}{inline}{}
\usepackage{placeins} 
\usepackage{booktabs}
\usepackage{xspace}
\usepackage{fullpage}

\usepackage[utf8]{inputenc}
\usepackage[hidelinks]{hyperref}
\usepackage{color}
\usepackage[version=4]{mhchem}
\usepackage{booktabs}

\usepackage{subcaption}
\usepackage{listings,xspace}
\usepackage{pdfpages}
\usepackage{verbatim}
\usepackage{url}
\usepackage{enumerate}
\usepackage{physics}
\usepackage{multirow}

\usepackage{tikz}
\usetikzlibrary{circuits.ee.IEC, positioning}
\graphicspath{{./gfx/}}

\usepackage[per-mode=symbol]{siunitx}
\usepackage{cleveref}
\crefname{equation}{Eq.}{Eqs.}
\crefname{section}{Sec.}{Secs.}
\crefname{table}{Tab.}{Tabs.}
\crefname{figure}{Fig.}{Figs.}
\crefname{subfigure}{Fig.}{Figs.}

\usepackage[numbers]{natbib}
\usepackage{caption}
\newcommand{\lhto}{\ce{LH2}\xspace}

\journal{arXiv}

\begin{document}

\begin{frontmatter}

\title{Local Area Cooling versus Broad Area Cooling for Boil-Off Reduction in Large-Scale Liquid Hydrogen Storage Tanks}

\affiliation[NTNU_IMF]{
  organization={Department of Mathematical Sciences, Norwegian University of Science and Technology},
  city={Trondheim},
  country={Norway}}
\affiliation[SINTEF]{
  organization={SINTEF Energy Research},
  city={Trondheim},
  country={Norway}}
\author[NTNU_IMF,SINTEF]{Sindre Stenen Blakseth}
\ead{sindre.blakseth@sintef.no}
\author[SINTEF]{Ailo Aasen}
\author[NTNU_IMF]{André Massing}
\author[SINTEF]{Petter Nekså}

\date{\today}

\begin{abstract}
  Future use of liquid hydrogen (\lhto) as an effective energy carrier will  
  require elimination or minimization of hydrogen boil-off that is not utilised by demands in the value chain.  
  The present work promotes local area cooling (LAC) 
  as a promising boil-off reduction technology. 
  In contrast to the more conventional broad area cooling (BAC), 
  LAC targets local, concentrated heat flows e.g.\ through tank support structures. 
  This yields important practical benefits, especially for large-scale tanks, 
  due to the order-of-magnitude reduction in the size of the cooling system. 
  Such benefits include lower capital costs and simpler installation, maintenance and coolant management. 
  LAC applied outside the outer tank wall is particularly attractive for tanks with evacuated insulation. 
  
  In a series of numerical studies, we use the finite element method to 
  evaluate the thermal performance of LAC and BAC in the context of ship-borne \lhto transport. 
  The studies concern \SI{40000}{\meter^3}-capacity, skirt-supported tanks insulated using evacuated perlite or helium-filled polyurethane (HePUR) foam. 
  For the perlite-insulated tank, LAC and BAC with liquid nitrogen coolant can reduce 
  the daily boil-off rate from 0.04\%/day to, respectively, 0.011\%/day and 0.004\%/day. 
  The corresponding numbers for CO$_2$-based refrigeration are 0.031\%/day and 0.028\%/day. 
  For the HePUR-insulated tank, which has a higher baseline boil-off rate of 0.24\%/day, 
  reduced boil-off rates down to 0.17\%/day and 0.04\%/day are achievable using LAC and BAC, respectively. 
  LAC and BAC both offer increased power efficiency in comparison to reliquefaction only. 
\end{abstract}


\end{frontmatter}

\section{Introduction}
\label{sec:introduction}

\subsection{Motivation}
\label{sec:motivation}

Hydrogen has been highlighted as a promising energy carrier in the global transition from fossil fuels to renewable energy sources.
However, its efficacy is contingent on the development of efficient, large-scale storage and transport solutions.
One of the main issues in this matter is to achieve satisfactory energy density, especially during transport.
Numerous solutions have been proposed to increase the energy density of hydrogen during transport, including compression~\cite{rivard2019hsf}, liquefaction~\cite{alghafri2022hla}, and various forms of chemical storage, e.g.\ via ammonia~\cite{giddey2017aaa} or liquid organic hydrogen carriers~\cite{preuster2017loh}, and physisorption on metal-organic frameworks~\cite{park20222kh}.
In the present work, our focus will be on liquid hydrogen (\lhto), which has been highlighted as a viable solution for ship-borne transport of large volumes over long distances~\cite{ishimoto2020lpa}.

The advantages of \lhto include high purity, high gravimetric energy density, and possibility of storage at ambient pressure.
These advantages are offset by the high energy demand of the liquefaction process, as well as hydrogen's low normal boiling point (\SI{20}{\kelvin}) and its low volumetric heat of evaporation.
The latter properties imply that \lhto storage vessels must be carefully designed in order to avoid excess formation of boil-off gas (BOG) due to heat ingress. 
Crucially, the insulation concepts that have previously been used for liquid natural gas (LNG) carrier ships are not satisfactory for transport and storage of \lhto.
Hence, \lhto tanks are customarily constructed using specialized, often evacuated, high-performance insulation, e.g.\ based on perlite~\cite{krenn2012doa}, glass bubbles~\cite{fesmire2022eel}, or multilayer insulation~\cite{muragishi2021htd}.

Depending on the application, the use of high-performance insulation materials may, by itself, limit heat ingress to acceptable levels.
Indeed, existing \lhto tanks, such as those operated by NASA and Kawasaki Heavy Industries~\cite{krenn2012doa, muragishi2021htd}, have largely been constructed without additional cooling measures.
However, with this approach, some BOG will necessarily be generated.\footnote{The review by \citet{morales-ospino2023str} gives a great overview of the mechanisms causing boil-off and some basic BOG handling concepts.}
If the BOG rate exceeds the demand of any related hydrogen-powered systems,
such as a ship's propulsion system, the excess BOG must be accommodated somehow.
Any excess BOG may be contained within the tank, which must then be designed to handle the resulting pressure increase,
or the BOG will have to be released from the tank and either vented, stored in an auxiliary tank, or reliquefied.\footnote{Or, more esoterically, absorbed using metal hydrides, as suggested by \citet{rosso1987col}.}
For ship-borne transport, these solutions are all associated with significant drawbacks.
For the tank sizes in question, self-pressurization would require prohibitively thick tank walls.
Moreover, venting constitutes loss of cargo, which has an additional drawback that it contributes towards increased green house effect~\cite{sand2023ama}.
Reliquefaction or auxiliary storage are arguably more feasible options,
but still necessitate more complicated hydrogen handling systems, increased on-ship area usage, and, especially in the case of reliquefaction, further technological development in order to meet weight, space and capacity requirements.
Due to these considerations, there are great potential benefits from reducing BOG generation.

\subsection{Related Work}
\label{sec:related_work}

BOG generation can be reduced in two principal ways:
Measures can be taken to reduce heat ingress from the environment,
or cooling can be supplied directly to the contained hydrogen in order to counteract heat ingress.
The latter approach, which we refer to as direct hydrogen cooling (DHC), can be realized in many ways.
Examples include injection of sub-cooled \lhto~\cite{liu2016pso, kartuzova2014sas, ho2008nid},
circulation of sub-cooled \lhto in a heat pipe~\cite{zakar2015clh},
circulation of helium in submerged cooling lines~\cite{baik2015tpp},
expansion cooling using Joule--Thompson valves~\cite{zhou2023eio},
catalysis of para-ortho conversion~\cite{matveev2023mol}, or
placing the cold head of a cryocooler inside the tank~\cite{ho2008taf, wan2023nso}.
In addition to reducing\slash eliminating boil-off, DHC can serve a dual purpose of managing the pressure and fluid flow within the tank, which is especially important in microgravity.
Hence, DHC has received significant attention in aerospace applications, including NASA's long-standing efforts to develop space- and weight-efficient zero boil-off concepts for propellant tanks~\cite{hastings2001aoo, plachta2018nct}.
However, the need to supply cooling at temperatures below the normal boiling point of \lhto is a significant drawback.
Possibly due to the limited capacity of existing cryocoolers operating at sub-\lhto temperatures, DHC approaches have mainly been considered for smaller tanks.
When instead considering reduction of heat ingress from the environment,
one frequently studied approach is to install a cooled metal sheet encompassing the tank.
If the metal sheet (aka.\ shield) is cooled using BOG, it is referred to as a vapor-cooled shield (VCS).
VCS has been considered for liquid helium and liquid hydrogen storage at least since the 1960s~\cite{paivanas1965maa}.
Typically, VCS cooling is achieved by passing BOG through one or more tubes that are in thermal contact with the shield.
Already in 1970, \citet{davies1970hta} used a lumped-parameter model to study the thermodynamically optimal placement and sizing of such tubes.
Tube optimization can also be carried out by means of finite volume simulations, as performed by~\citet{zhu2024cdo} for a 12.56-liter cylindrical \lhto tank.
Another aspect to consider is the optimal placement of the shield itself inside the insulation space. \citet{hofmann2004ttb} suggests a set of lumped-parameter models for this purpose.
In terms of performance, recent computational studies suggest that a single VCS can reduce total heat ingress by 50--70\%, depending on factors such as tank geometry and choice of insulation material~\cite{sun2023aqt, liu2023dao, jiang2021ttb, zheng2020anc, zheng2019tma, jiang2018coo}.
These estimates are more optimistic than the 25--35\% reduction measured by \citet{liggett1993slp}.

The effectiveness of VCS can be increased by installing multiple shields~\cite{yu2023dao}.
Connecting the shields in series appears to be favorable in comparison to a parallel setup~\cite{kim2000tda}.
\citet{yu2023dao} estimate that a dual, series-connected VCS can reduce heat ingress by more than 80\% for a 4000-\si{\meter^3} spherical \lhto tank.
Further improvements can be achieved by combining VCS with a para-ortho converter.
Compared to a benchmark VCS without para-ortho conversion, the introduction of the converter appears to reduce heat ingress by 5--12\%~\cite{meng2023cea, shi2023pao, lv2024tao, xu2023pco}.

VCS is included in a broader category of cooling technologies called broad area cooling (BAC).
In addition to VCS, BAC also includes cooling arrangements where shields are cooled using external cooling sources.
Unlike VCS, these other technologies are \emph{active}, meaning they require external power input.
Hence, they are less thermodynamically efficient.
However, they offer greater flexibility in terms of, e.g., coolant temperature, mass flow rate, and maximum cooling effect, while also allowing a simpler BOG handling system.
A broad selection of BAC cooling sources has been considered in the literature,
including pulse-tube cryocoolers~\cite{zheng2020too},
refrigeration units powered by a BOG-driven fuel cell~\cite{zheng2019ani, xu2020ahl},
and reverse turbo-Brayton cycles~\cite{plachta2016zbs}.
For installations with multiple shields, dual-stage cryocoolers can also be considered~\cite{segado2010ccd}.
Cooling can either be supplied through a network of coolant-filled tubes~\cite{plachta2008cpb, feller2010coa} or localized ``cold sectors''~\cite{posada2005aso}.
Cooling loops can also be installed directly on the tank wall~\cite{haberbusch2010don}.

Unlike passive VCS, active BAC technologies have the potential to achieve zero boil-off, as verified both numerically~\cite{haberbusch2004toz} and experimentally (for liquid nitrogen)~\cite{plachta2016zbs}.
However, this either requires cooling the shield to the temperature of the stored cryogen, or a tank that can handle internal pressure increase.
Hence, BAC is more commonly applied to reduce, rather than eliminate, BOG generation.
For example, \citet{plachta2014cbr} and \citet{johnson2014tat} report experimental heat ingress reductions in the range of 50--60\% for a \SI{1.2}{\meter}-diameter, cylindrical \lhto tank.
\citet{feller2011doa} reports even larger reductions, up to 80\%, for a 500-liter liquid nitrogen tank.
Heat ingress reduction around 80\% is also reported in the numerical study by~\cite{xu2020ahl}.
In any case, it is important to balance the power consumption of the BAC with the associated savings in BOG handling, e.g. related to reliquefaction.
\citet{chen2017rot} and \citet{zheng2020too} find that BAC reduces total power requirements by, respectively, 88\% and 70\%, when comparing to reliquefaction of the BOG from tanks without BAC.

As an alternative to BAC, one could consider more localized cooling in regions of the containment system where the heat flux is large.
We refer to this as \emph{local area cooling} (LAC).
Naturally, a reduction in the area where cooling is applied will necessarily reduce the amount of heat that can be intercepted by the cooling system.
However, if a significant fraction of the heat enters the tank through localized paths, such as through support structures or pipelines, localized cooling may still be highly effective.
LAC may also have certain practical benefits in comparison to BAC.
For example, it does not require the installation of a potentially heavy and expensive shield,
its required coolant pipeline network would be shorter and hence easier to manage and less vulnerable to leaks,
and it would interfere less with regulatory inspections of the insulation space.
These benefits are especially pronounced for large tanks with diameters on the order of tens of meters, as is being considered e.g.\ for \lhto carrier ships.
A recent computational study \cite{aasen2024tpe} also found that approximately 70\% of the heat ingress of a large-scale, skirt-supported ship tank enters through the support skirt, indicating that such tanks may be prime candidates for localized cooling.
Yet, to the authors' best knowledge, LAC has not previously been studied for this application.

Indeed, even though localized cooling is a simple and known concept (e.g.\ being part of NASA's zero boil-off endeavours~\cite{plachta2008cpb, meyer2024npm}),
quantitative studies of its performance are scarce in the literature on \lhto storage.
Important exceptions are the works by \citet{kim2000tda}, \citet{plachta2014cbr}, and \citet{muratov2011iol}.
\citet{kim2000tda} find that applying localized cooling to the filling tube of an \lhto tank can reduce the amount of heat entering the tank through the filling tube by 55\%.
In the study by \citet{plachta2014cbr}, the main focus is on BAC, but the authors note that placing the tank support struts in contact with the BAC shield using aluminium straps can reduce the heat transfer through the struts by up to 68\%.
Finally, \citet{muratov2011iol} present a series of analytic considerations regarding the sizing of cooling tubes attached to tank support struts and filled with BOG from the tank.
Nonetheless, the opportunities of LAC remain largely unexplored.





\subsection{Present Contribution}
\label{sec:contribution}

The main contribution of the present work is a quantitative performance analysis of
local area cooling (LAC) for cryogenic storage tanks.
In this analysis, which also includes broad area cooling (BAC) as a contrasting alternative,
the finite element method is used to
model heat transfer within large-scale storage tanks for ship-borne \lhto transport.
The performance of LAC and BAC is compared along two axes: maximum boil-off rate reduction, and minimum power consumption to achieve some specified net boil-off rate.
For the latter, reliquefaction is included in cases where the active cooling does not enable sufficient boil-off reduction on its own.\footnote{We have also considered direct cooling of the contained \lhto as an alternative to reliquefaction, but this was found to be a less favorable option. Hence, direct hydrogen cooling is only considered in \ref{app:extended_results} and \ref{app:efficiency_degradation}.}
In addition to the quantitative thermal performance analysis, we also discuss other important practical matters pertaining to active cooling of large-scale, ship-borne \lhto tanks.

\subsection{Outline}
\label{sec:outline}

The \lhto containment system studied in the present work is described in Section~\ref{sec:system}.
The design geometry and material choices is outlined in Section~\ref{sec:geometry_and_materials},
while details regarding the LAC and BAC cooling systems are presented in Section~\ref{sec:active_cooling}.
Section~\ref{sec:modelling} concerns the mathematical models and numerical solutions techniques used to analyze the performance of LAC and BAC.
Heat transfer modelling, including boil-off estimation, is described in Section~\ref{sec:modelling_heat}, while power consumption estimation is covered in Section~\ref{sec:modelling_power}.
Performance optimization results are presented and discussed in Section~\ref{sec:results}.
The discussion continues in Section~\ref{sec:discussion} with practical aspects of active cooling and interesting avenues for future work.
Finally, concluding remarks are given in Section~\ref{sec:conclusions}.

\section{Containment System Description}
\label{sec:system}

\subsection{Design Geometry and Materials}
\label{sec:geometry_and_materials}

The \lhto containment system studied herein is a spherical, double-walled, skirt-support tank with an inner radius of \SI{21.2}{\meter}.
It is similar to the designs considered in \cite{aasen2024tpe} and in the patent application \cite{skogan2022lgs} by Moss Maritime.
We consider two variations of the containment system, distinguished only by the choice of primary insulation material.
One variation is insulated with perlite, which offers high insulation performance but requires high vacuum.
The alternative is helium-filled polyurethane (HePUR) foam insulation,
which is less performant but can be used at ambient pressure.

\begin{figure}
  \centering
  \includegraphics[width=0.45\textwidth]{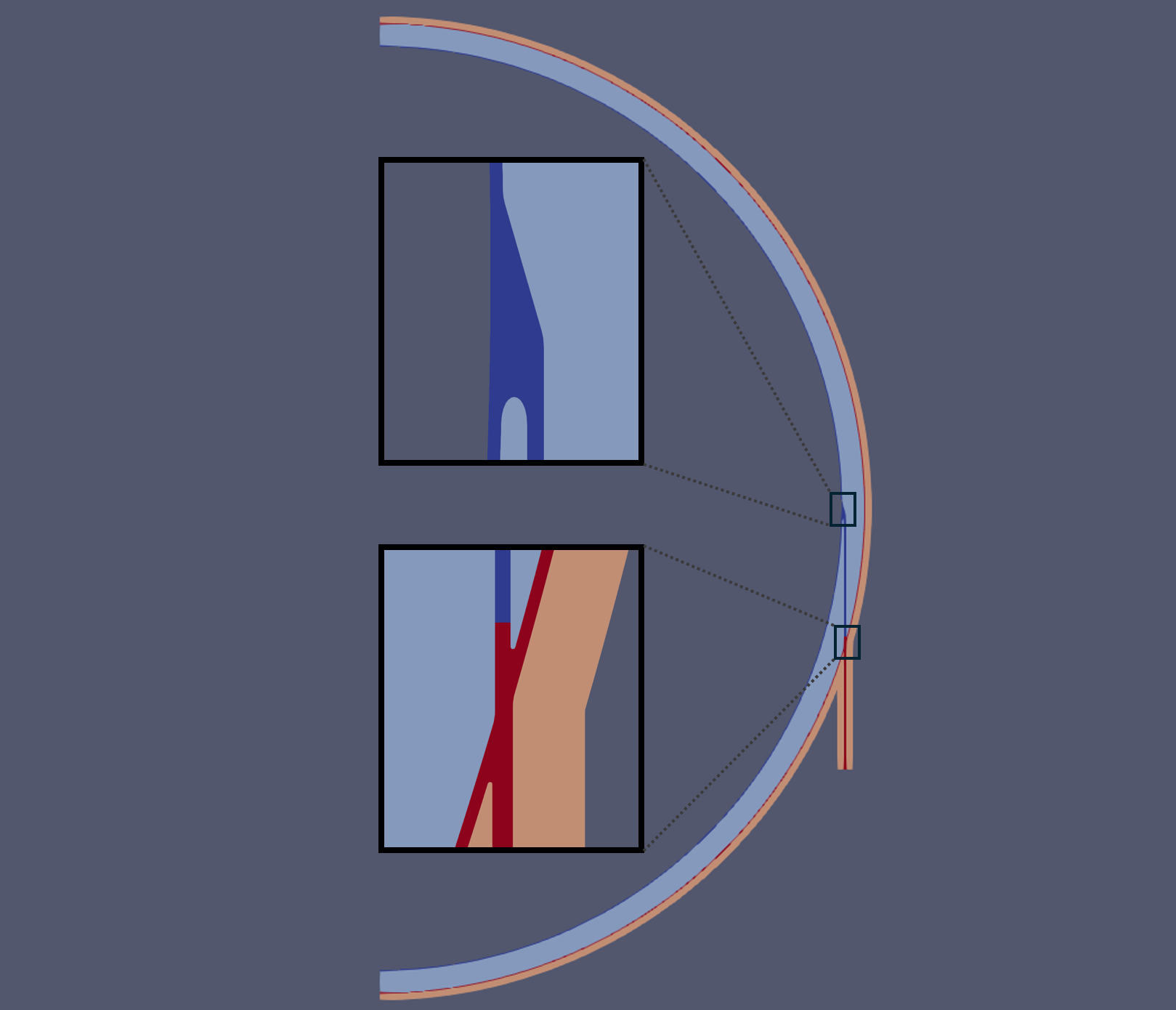}
  \caption{
    The \lhto containment system geometry,
    including detailed views of the ring mount (top infield) and the cross mount (bottom infield). Dark blue: 316 stainless steel, dark red: carbon steel, light blue: primary insulation material (perlite or He-filled polyurethane foam), orange: regular polyurethane foam.}
  \label{fig:geo}
\end{figure}

The containment system is illustrated in Figure~\ref{fig:geo}, and consists of the following main components:
\begin{enumerate}
  \item The primary tank wall, within which the \lhto is contained. Thickness: \SI{0.05}{\meter}.
  \item The secondary tank wall, which provides a secondary barrier against \lhto leakage, and also enables the use of vacuum insulation. Thickness: \SI{0.05}{\meter}.
  \item The primary insulation, which is placed between the primary and secondary tank walls. Thickness: \SI{1.0}{\meter}.
  \item The secondary insulation, which is placed outside the secondary tank wall. Thickness: \SI{0.3}{\meter}.
  \item The primary support skirt, which supports the primary tank wall. Thickness: \SI{0.065}{\meter}.
  \item The secondary support skirt, which connects to the inner hull of the ship and supports the entire containment system. Thickness: \SI{0.085}{\meter}.
  \item The ring mount, which connects the primary tank wall to the primary support skirt.
  \item The cross mount, which connects the primary support skirt and the secondary tank wall to the secondary support skirt.
\end{enumerate}
In Figure~\ref{fig:geo}, pale blue represents the primary insulation material (perlite or HePUR foam),
dark blue represents stainless steel 316, dark red represents carbon steel, and orange represents regular polyurethane foam.
Each support skirt segment is approximately \SI{5.3}{\meter} long.

\begin{figure}
  \centering
  \includegraphics[width=0.45\textwidth]{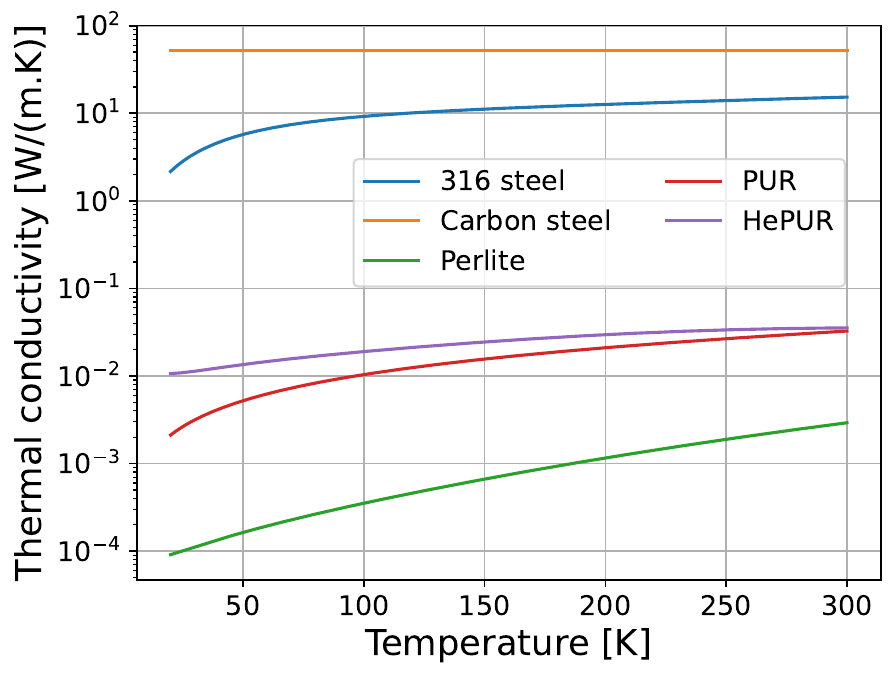}
  \caption{Thermal conductivities of the structural materials and insulation materials used in the present work. He-filled polyurethane (HePUR) foam has a somewhat higher thermal conductivity than regular polyurethane (PUR) foam, but is still preferred as a primary insulation material because the helium gas does not freeze out. Evacuated perlite offers the premier thermal insulation performance, but requires high vacuum, which can be challenging to maintain for tanks as large as those considered herein.
  Data sources: perlite and regular PUR~\cite{ratnakar2023etc}, carbon steel~\cite{cristea2020stc}, stainless steel and HePUR~\cite{nist2021ctr}.}
  \label{fig:k}
\end{figure}

Figure~\ref{fig:k} displays the thermal conductivites of the aforementioned materials as functions of temperature.
The thermal conductivity of the perlite insulation is evidently lower than that of the PUR foam.
This implies that the perlite-insulated tank will have less heat ingress, and thus also less boil-off, than the HePUR-insulated tank.
Moreover, for the former, a relatively larger portion of the total heat ingress will be transmitted through the support skirt, because the skirt's thermal conductivity will be higher in comparison to that of the primary insulation material.
As such, the two primary insulation choices enable investigation of how differences in passive thermal performance and heat flow patterns affect the optimal configuration and maximum efficacy of the active cooling systems under consideration.

\subsection{Active Cooling Systems}
\label{sec:active_cooling}

As mentioned in Section~\ref{sec:contribution}, we study three different active cooling technologies:
local area cooling (LAC), broad area cooling (BAC), and reliquefaction.
Below, we provide the details needed to incorporate these technologies into our models.

\paragraph{Local area cooling} The LAC considered herein is a liquid-filled tube running along the outer circumference of the tank's support skirt.
The tube may be installed on either the primary or the secondary support skirt. The liquid is assumed to hold a temperature $T_{\mathrm{cool},\textsc{lac}}$ in the range $[\SI{77}{\kelvin}, \SI{285}{\kelvin}]$.

\paragraph{Broad area cooling} We consider a single, spherical BAC shield installed within the primary insulation space (i.e., between the two tank walls) and attached to the primary support skirt.
The shield is assumed to hold a uniform temperature $T_{\mathrm{cool},\textsc{bac}}$ in the range $[\SI{77}{\kelvin}, \SI{285}{\kelvin}]$.

\paragraph{Reliquefaction} In the present work, any excess BOG from the tank is handled by reliquefaction.
The reliquefaction unit is assumed to not provide any sub-cooling, so that the
thermal equilibrium within the tank remains undisturbed when reliquefied hydrogen is 
pumped back into the tank. We also assume that the capacity of the reliquefaction unit is not a limiting factor.
For more detailed considerations of reliquefaction units, we refer to the work by \citet{kim2023hrp}.

\paragraph{System configurations} We combine the technologies above to create six different system configurations,
as listed in Table~\ref{tab:configs}.
Configurations~4--6 can achieve zero BOG due to the inclusion of reliquefaction.

\begin{table}
  \centering
  \begin{tabular}{llcccc}
    \toprule
    \verb|#| & Configuration & LAC & BAC & Reliq. \\
    \midrule
    1 & Passive & No & No & No \\
    2 & LAC only & Yes & No & No \\
    3 & BAC only & No & Yes & No \\
    4 & Reliq.\ only & No & No & Yes \\
    5 & LAC + Reliq. & Yes & No & Yes \\
    6 & BAC + Reliq. & No & Yes & Yes \\
    \bottomrule
  \end{tabular}
  \caption{The cooling system configurations considered in the present work.}
  \label{tab:configs}
\end{table}

\section{Numerical Modelling}
\label{sec:modelling}

\subsection{Heat Transfer Modelling}
\label{sec:modelling_heat}

\subsubsection{Estimating Temperatures}
\label{sec:modelling_temperatures}

Under steady state conditions, the temperature field $T$ within the \lhto containment system is given by the steady heat equation
\begin{equation}
  -\nabla \cdot (k(T) \nabla T) = f,
\end{equation}
where $k$ denotes thermal conductivity and $f$ is a source term representing internal heat generation\slash extraction.
This equation must be supplemented by a set of boundary conditions in order to have a unique solution.
On the inner surface of the inner tank wall, denoted $\Gamma_{\mathrm{in}}$,
we set a fixed temperature $T = \SI{21}{\kelvin}$, representing the saturation temperature of \lhto at \SI{1.2}{\bar}.
In other words, we assume that the tank is approximately full, and that $\Gamma_{\mathrm{in}}$ is in perfect thermal contact with the contained \lhto.
On the outer surface of the secondary insulation and the bottom surface of the secondary skirt, collectively denoted $\Gamma_{\mathrm{out}}$, we set $T = \SI{300}{\kelvin}$, representing a typical ambient temperature.

For a containment system without active cooling, the above description is sufficient.
To include active cooling, two different approaches may be taken.
If a given cooling power $P_{\mathrm{cool}}$ is prescribed in a given region with volume $V$,
this can be accounted for by setting $f = -P_{\mathrm{cool}}/V$ in that region and $f=0$ elsewhere.
Alternatively, since we assume the cooling arrangement to be isothermal and in perfect thermal contact with its surroundings,
we can simply ``cut out'' the cooling arrangement from the computational domain and prescribe $T = T_{\mathrm{cool}}$ on the boundary section $\Gamma_{\mathrm{cool}}$ that then appears.
Here, $T_{\mathrm{cool}}$ is the coolant temperature, which is a free design parameter.
In this case, $f=0$ everywhere in the computational domain.
Since the latter approach facilitates a domain triangulation with fewer elements than the former, it is more computationally efficient.
For that reason, the latter approach will be used herein.

To summarize, our complete mathematical model is
\begin{alignat}{3}
  \label{eq:heat_eq_start}
  -\nabla \cdot (k(T) \nabla T) &= 0, \quad &&\mathrm{in}\ &&\Omega, \\
  T &= \SI{21}{\kelvin}, \quad &&\mathrm{on}\ &&\Gamma_{\mathrm{in}}, \\
  T &= \SI{300}{\kelvin}, \quad &&\mathrm{on}\ &&\Gamma_{\mathrm{out}}, \\
  \label{eq:heat_eq_end}
  T &= T_{\mathrm{cool}}, \quad &&\mathrm{on}\ &&\Gamma_{\mathrm{cool}},
\end{alignat}
where $\Omega$ is the interior of the computational domain.
For the passive configuration~1, $\Gamma_{\mathrm{cool}} = \emptyset$.
We also highlight that $k(T)$ is discontinuous at material interfaces, since the different materials have different thermal conductivities (cf.\ Figure~\ref{fig:k}).

Equations~\eqref{eq:heat_eq_start}--\eqref{eq:heat_eq_end} can be solved numerically using the finite element method.
To this end, we use the Python packages Netgen\slash NGSolve~\cite{schoberl1997naa, schoberl2014c1i} and follow the same procedure as in~\cite{aasen2024tpe}
to solve the equations on a two-dimensional mesh by exploiting the system's rotational symmetry.
The use of a 2D mesh creates an additional boundary segment $\Gamma_{\mathrm{axis}}$ in the computational domain.
$\Gamma_{\mathrm{axis}}$ coincides with the axis of symmetry and requires a homogeneous Neumann (i.e., zero heat flux) boundary condition for energy to be conserved.

\subsubsection{Estimating Heat Flows}
\label{sec:modelling_heat_flows}

The temperature field within the containment system is, by itself, not so interesting for the purposes of the present work.
More interesting are the heat flows across the various system surfaces,
since knowing these would allow us to estimate the boil-off mass flow from the tank and the power consumption of the active cooling systems.
Consquently, given the estimated temperature field $T_h$, we want estimates $Q_{\mathrm{in}}$, $Q_{\mathrm{out}}$ and $Q_{\mathrm{cool}}$ of the heat flows across $\Gamma_{\mathrm{in}}$, $\Gamma_{\mathrm{out}}$ and $\Gamma_{\mathrm{cool}}$.
For this, we make use of the relation
\begin{equation}
  Q_{\mathrm{in}} = -\int\limits_{\Omega} k(T_h) \nabla T_h \cdot \nabla v_{\mathrm{in}} \, \mathrm{d}\Omega,
  \label{eq:Q_in}
\end{equation}
where $v_{\mathrm{in}}$ is some continuous, piece-wise differentiable function that satisfies
\begin{equation}
  v_{\mathrm{in}} =
  \begin{cases}
    1, &\mathrm{on}\ \Gamma_{\mathrm{in}}, \\
    0, &\mathrm{elsewhere\ on\ the\ boundary}.
  \end{cases}
\end{equation}
We choose $v_{\mathrm{in}}$ to be the piecewise linear function that is equal to 1 at all mesh nodes on $\Gamma_{\mathrm{in}}$ and 0 on all other mesh nodes.

Equation~\eqref{eq:Q_in} can be derived by multiplying Equation~\eqref{eq:heat_eq_start} with $v_{\mathrm{in}}$, integrating over $\Omega$, and applying partial integration.
An analogous relation can be derived for $Q_{\mathrm{out}}$, while we utilize global energy conservation (which is preserved by estimates of the form in Equation~\eqref{eq:Q_in}) to estimate $Q_{\mathrm{cool}}$ as
\begin{equation}
  Q_{\mathrm{cool}} = -(Q_{\mathrm{in}} + Q_{\mathrm{out}}).
  \label{eq:Q_cool}
\end{equation}

\subsubsection{Estimating Boil-Off}
\label{sec:modelling_boil-off}

We assume that all the heat entering the tank through $\Gamma_{\mathrm{in}}$ contributes to evaporation of the contained \lhto.
Hence, the boil-off mass flow rate $\dot{m}_{\textsc{bog}}$ from the tank can be estimated as
\begin{equation}
  \dot{m}_{\textsc{bog}} = \frac{Q_{\mathrm{in}}}{\Delta H_{\mathrm{vap}}},
  \label{eq:boil-off}
\end{equation}
where $\Delta H_{\mathrm{vap}} = \SI{0.45}{\mega\joule\per\kilogram}$ 
is the specific latent heat of vaporization of hydrogen.

\subsection{Estimating Power Consumption}
\label{sec:modelling_power}

In order to estimate the power consumption from each active cooling configuration listed in Table~\ref{tab:configs},
we must first estimate the individual power consumptions of the LAC, BAC and reliquefaction systems.
In this section, we describe how to obtain such estimates.

In general, the power consumption $P$ of a cooler is characterized by its coefficient of performance (COP) via the relation
\begin{equation}
  P = Q / \mathrm{COP},
\end{equation}
where $Q$ is the amount of heat removed by the cooler.
If the cooler removes heat at a temperature $T_{\textsc{c}}$ and dumps it at a temperature $T_{\textsc{h}}$,
its COP is given by
\begin{equation}
  \mathrm{COP} = \eta\frac{T_{\textsc{c}}}{T_{\textsc{h}} - T_{\textsc{c}}},
\end{equation}
where $\eta$ is the cooler's Carnot efficiency.\footnote{By ``Carnot efficiency'', also called ``second-law efficiency'', we mean the ratio between the COP of a real cycle and the theoretical COP of the corresponding ideal cycle.
This is not to be confused with the ``Carnot cycle efficiency'', which is the efficiency of the Carnot cycle.}

The power consumption of the LAC and BAC systems will be dominated by the power consumption of their respective refrigeration units.
These remove the heat $Q_{\mathrm{cool}}$, given by Equation~\eqref{eq:Q_cool}, at the temperature $T_{\mathrm{cool}}$ and dump it to the ambient temperature $T_{\mathrm{amb}}$.
Consequently, the power consumptions of LAC and BAC are estimated as
\begin{align}
  P_{\textsc{lac}} &= \frac{Q_{\mathrm{cool}}}{\eta_{\textsc{lac}}\frac{T_{\mathrm{cool},\textsc{lac}}}{T_{\mathrm{amb}} - T_{\mathrm{cool}}}}, \\
  P_{\textsc{bac}} &= \frac{Q_{\mathrm{cool}}}{\eta_{\textsc{bac}}\frac{T_{\mathrm{cool},\textsc{bac}}}{T_{\mathrm{amb}} - T_{\mathrm{cool}}}}.
\end{align}
We assume the following Carnot efficiencies: $\eta_{\textsc{lac}} = \eta_{\textsc{bac}} = 0.5$~\cite{zuhlsdorf2023hhp}.\footnote{In practice, the Carnot efficiencies may degrade as $T_{\mathrm{cool}}$ decreases.
The implications of such efficiency degradation are discussed in \ref{app:efficiency_degradation}.
Moreover, we have verified that calculating the COP as $\eta(T_{\mathrm{cool}} - 2)/(T_{\mathrm{amb}}+5 - (T_{\mathrm{cool}} - 2))$, in order
to account for the finite temperature jumps present in real systems,
does not alter any conclusions drawn in the present work.
This justifies the assumption of perfect thermal conctact made in Section~\ref{sec:modelling_temperatures}.}




The power consumption $P_{\mathrm{reliq}}$ of a reliquefaction unit can be calculated as
\begin{equation}
  P_{\mathrm{reliq}} = \chi \dot{m}_{\mathrm{reliq}},
  \label{eq:power_reliq}
\end{equation}
where $\chi = \SI{5}{\kWh\per\kilogram}$ is the energy required to reliquefy one kilogram of hydrogen~\cite{kim2023hrp},
and $\dot{m}_{\mathrm{reliq}}$ is the mass flow of hydrogen to be reliquefied.
In general, the objective of the reliquefaction unit is to reliquefy any \emph{excess} boil-off, meaning that
\begin{equation}
  \dot{m}_{\mathrm{reliq}} = \mathrm{max}(0, \dot{m}_{\textsc{bog}} - \dot{m}_{\mathrm{fuel}}),
\end{equation}
where $\dot{m}_{\textsc{bog}}$ is the boil-off mass flow (given by Equation~\eqref{eq:boil-off}),
and $\dot{m}_{\mathrm{fuel}}$ is the demand of any hydrogen-powered systems fuelled from the tank (which may be zero).
In practice, $\dot{m}_{\mathrm{fuel}}$ will depend greatly on parameters such as choice of propulsion system, ship design and weather conditions.
Instead of making specific and limiting assumptions about these parameters, we simply assume
\begin{equation}
  \dot{m}_{\mathrm{reliq}} = (1 - \phi) \dot{m}_{\textsc{bog}, \mathrm{ref}},
\end{equation}
for some $\phi \in [0,1]$.
Here, $\dot{m}_{\textsc{bog}, \mathrm{ref}}$ is the boil-off mass flow from the passive baseline configuration~1.
Note that $\phi = 0$ corresponds to $\dot{m}_{\mathrm{fuel}} = 0$.
In Section~\ref{sec:min_power}, we consider $\phi=0$ and $\phi=0.5$.

\section{Results}
\label{sec:results}

The numerical results herein will be presented in three parts.
First, in Section~\ref{sec:baseline_results}, we establish the heat ingress and boil-off rate of the
perlite-insulated tank and the HePUR-insulated tank for the passive baseline configuration~1.
Thereafter, in Section~\ref{sec:min_boil-off}, we aim to identify
the minimum boil-off rate achievable using LAC (configuration~2) and BAC (configuration~3).
Finally, in Section~\ref{sec:min_power}, we estimate the minimum
power consumption required to achieve 50\% and 100\% net boil-off reduction
using reliquefaction with and without LAC/BAC (configurations~4--6).

\subsection{Baseline (No Active Cooling)}
\label{sec:baseline_results}

In this section, we present results for the baseline configuration~1,
which has no active cooling.
Using the methods presented in Section~\ref{sec:modelling_heat}
and a computational mesh with 131 824 elements,\footnote{The mesh resolution was determined from the grid refinement study described in \ref{app:grid_refinement}.}
we estimate that the Perlite-insulated tank has a baseline heat ingress of
$\SI{5905}{\watt}$.
Meanwhile, the HePUR-insulated tank
has an estimated baseline heat ingress of $\SI{35350}{\watt}$.
These heat ingress values correspond to boil-off rates of roughly 0.04\% per day and 0.24\% per day, respectively.

\begin{figure}
  \centering
  \begin{subfigure}[t]{0.23\textwidth}
    \includegraphics[width=\textwidth,trim={55cm 0 55cm 0},clip]{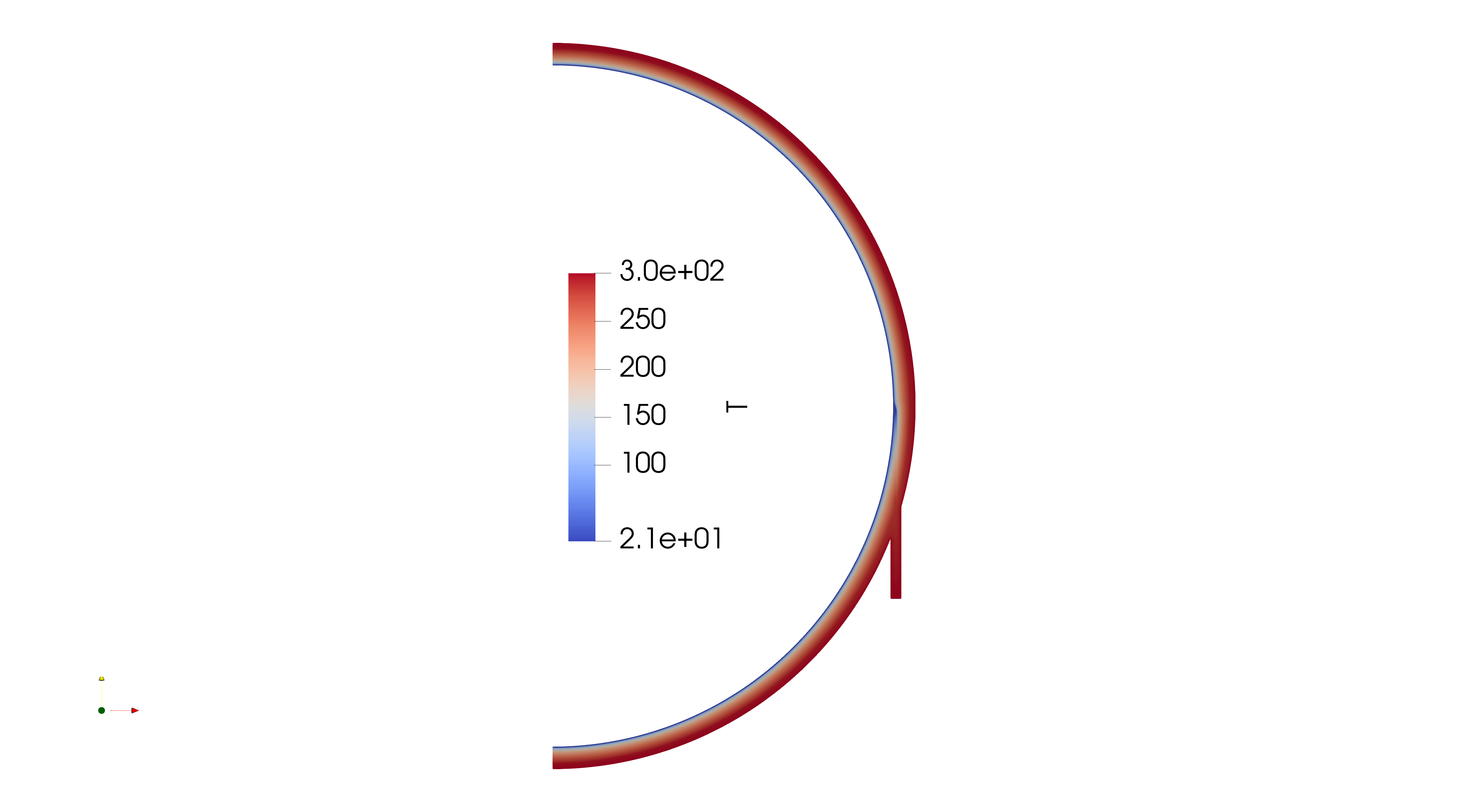}
    \caption{Full profile}
    \label{fig:baseline_temp_A_full}
  \end{subfigure}
  \begin{subfigure}[t]{0.23\textwidth}
    \includegraphics[width=\textwidth,trim={100cm 30cm 60cm 45cm},clip]{gfx/baseline_temp_A.png}
    \caption{Close-up of the area surrounding the support skirt.}
    \label{fig:baseline_temp_A_zoom}
  \end{subfigure}
  \caption{Temperature field in the perlite-insulated tank for the passive baseline Configuration~1.}
  \label{fig:baseline_temp_A}
\end{figure}

\begin{figure}
  \centering
  \begin{subfigure}[t]{0.23\textwidth}
    \includegraphics[width=\textwidth,trim={55cm 0 55cm 0},clip]{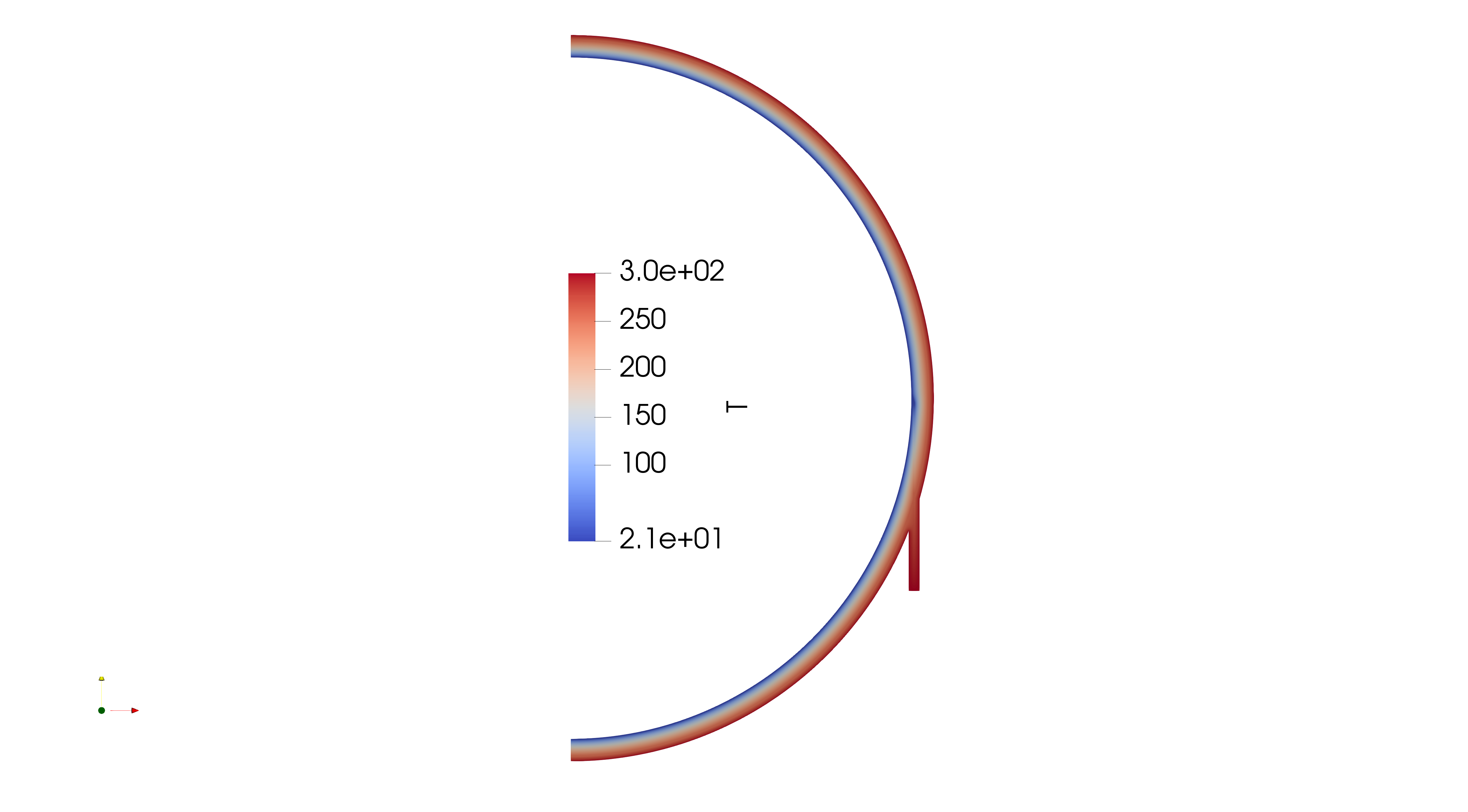}
    \caption{Full profile}
    \label{fig:baseline_temp_B_full}
  \end{subfigure}
  \begin{subfigure}[t]{0.23\textwidth}
    \includegraphics[width=\textwidth,trim={100cm 30cm 60cm 45cm},clip]{gfx/baseline_temp_B.png}
    \caption{Close-up of the area surrounding the support skirt.}
    \label{fig:baseline_temp_B_zoom}
  \end{subfigure}
  \caption{Temperature field in the HePUR-insulated tank for the passive baseline Configuration~1.}
  \label{fig:baseline_temp_B}
\end{figure}

The temperature field within the entire containment system is shown in Figures~\ref{fig:baseline_temp_A} and~\ref{fig:baseline_temp_B}
for perlite-insulated tank and the HePUR-insulated tank, respectively.
Since perlite is the more performant primary insulation material,
the perlite-insulated tank exhibits a comparatively larger temperature gradient within the primary insulation space.
In other words, the temperature in the vicinity of the secondary tank wall tends to be warmer for the perlite tank than for the HePUR-tank.
This observation has notable implications for optimal cooling system placements and coolant temperatures, as will be discussed in Section~\ref{sec:min_boil-off}.

\FloatBarrier

\subsection{Minimizing Boil-Off}
\label{sec:min_boil-off}

In this section, we analyze the potential of active cooling to decrease boil-off generation.
More specifically, for LAC and BAC (configurations~2 and~3 in Table~\ref{tab:configs}),
we analyze the impact of coolant temperature
and cooling shield\slash tube placement on the boil-off rate.
This analysis will be conducted without regard to the cooling system's power consumption,
which will be addressed in Section~\ref{sec:min_power}.



For BAC, we consider seven different shield positions, characterized by $\zeta \in \{0.25, 0.35, \dots, 0.85\}$,
where
\begin{equation}
  \zeta = \frac{r_{\mathrm{shield}} - R_{\mathrm{in}}}{\tau_{\mathrm{insul}}},
\end{equation}
$r_{\mathrm{shield}}$ is the inner radius of the BAC shield,
$R_{\mathrm{in}} = \SI{21.25}{\meter}$ is the outer radius of the inner tank wall,
and $\tau_{\mathrm{insul}} = \SI{1.0}{\meter}$ is the thickness of the primary insulation.

For LAC, we consider the analogous seven positions on the outer surface of the primary skirt,
as well as one position at the intersection between the secondary support skirt and the cross mount.
Thus, including the 7 BAC setups, we have a total of $7+7+1=15$ different cooling setups to consider for each choice of primary insulation.

\begin{figure}
  \centering
  \includegraphics[width=0.45\textwidth]{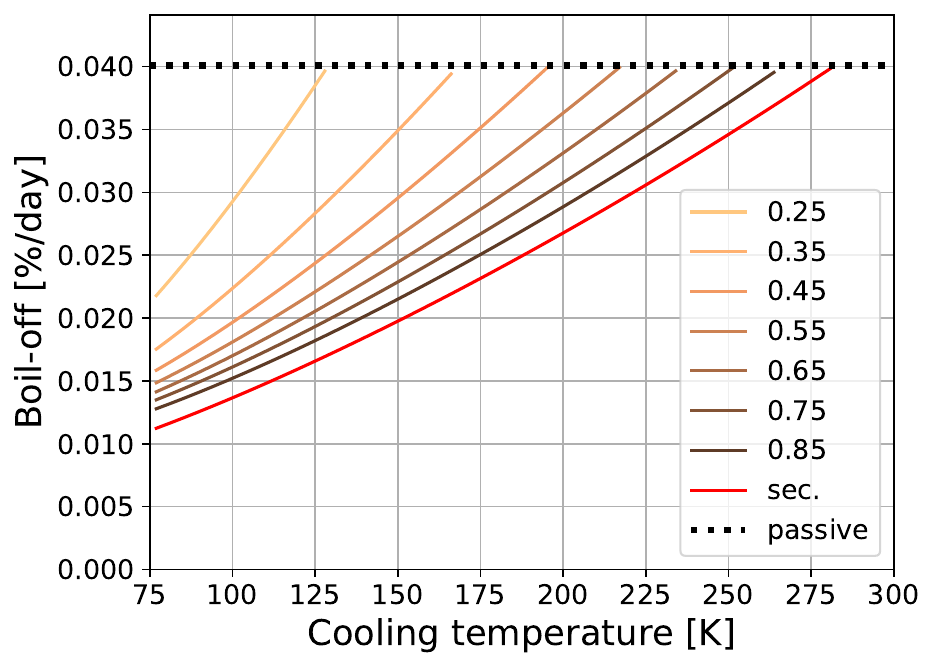}
  \caption{Boil-off rate as a function of cooling temperature for LAC (configuration~2) on the perlite tank.
  Red line: the cooling tube is placed on the secondary skirt.
  Remaining colored lines: different cooling tube placements along the primary skirt (the legend indicates their respective $\zeta$-values).
  Black line: the passive baseline configuration~1.}
  \label{fig:boil-off_LAC_A}
\end{figure}

\begin{figure}
  \centering
  \includegraphics[width=0.45\textwidth]{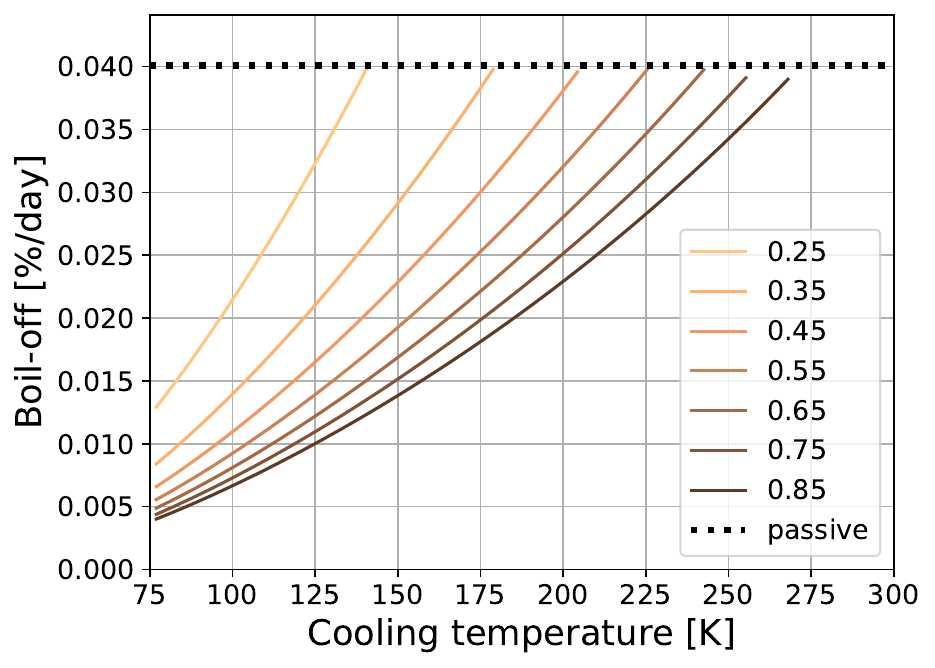}
  \caption{Boil-off rate as a function of cooling temperature for BAC (configuration~3) on the perlite tank.
  Colored lines: different cooling shield placements within the primary insulation space (the legend indicates their respective $\zeta$-values).
  Black line: the passive baseline configuration~1.}
  \label{fig:boil-off_BAC_A}
\end{figure}

\begin{figure}
  \centering
  \includegraphics[width=0.45\textwidth]{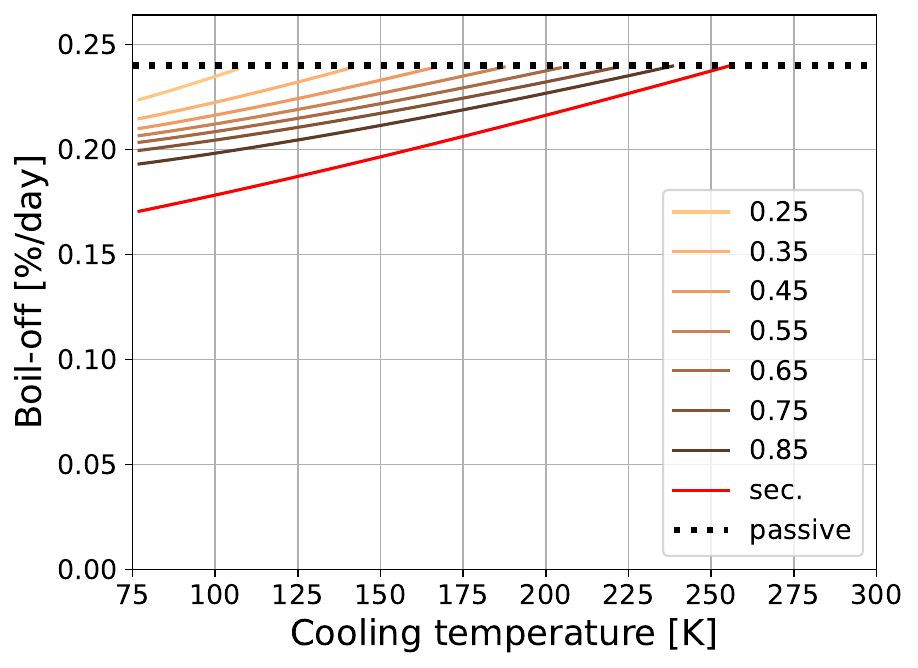}
  \caption{Boil-off rate as a function of cooling temperature for LAC (configuration~2) on the HePUR tank.
  The coloring scheme is the same as in Figure~\ref{fig:boil-off_LAC_A}.}
  \label{fig:boil-off_LAC_B}
\end{figure}

\begin{figure}
  \centering
  \includegraphics[width=0.45\textwidth]{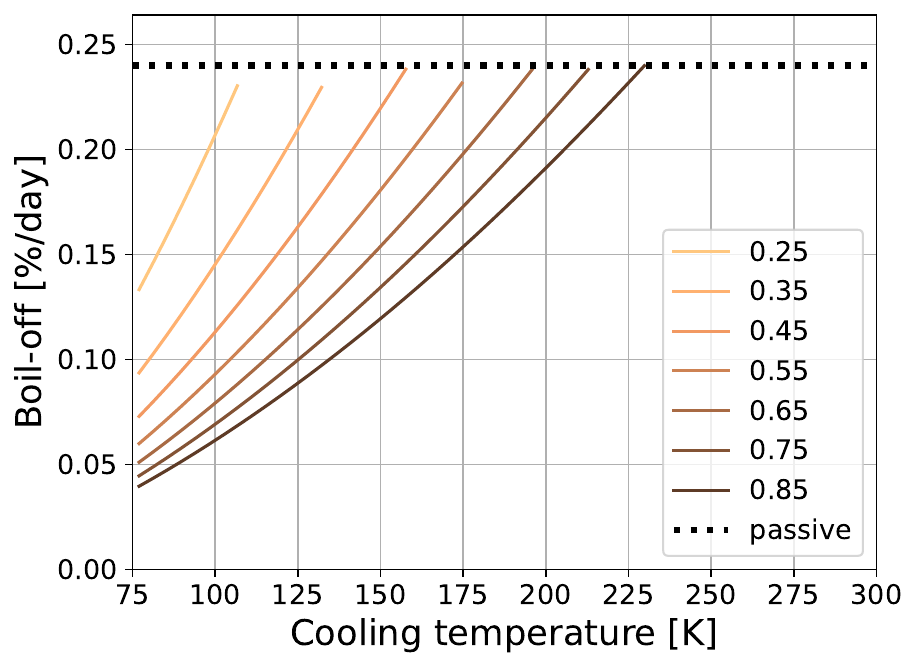}
  \caption{Boil-off rate as a function of cooling temperature for BAC (configuration~3) on the HePUR tank.
  The coloring scheme is the same as in Figure~\ref{fig:boil-off_LAC_B}.}
  \label{fig:boil-off_BAC_B}
\end{figure}

For each cooling setup and primary insulation choice, we compute the resulting boil-off rate
for 50 different coolant temperatures $T_{\mathrm{cool}}$ in the range $[\SI{77}{\kelvin}, \SI{285}{\kelvin}]$.
The resulting heat ingress profiles for the perlite tank are shown in Figures~\ref{fig:boil-off_LAC_A} and~\ref{fig:boil-off_BAC_A} for LAC and BAC, respectively.
The analogous profiles for the HePUR-tank are shown in Figures~\ref{fig:boil-off_LAC_B} and~\ref{fig:boil-off_BAC_B}.

Across all four figures, a couple of persistent trends can be observed.
First, for a fixed cooling location, reducing the coolant temperature will always reduce the boil-off rate.
Second, for a fixed coolant temperature, shifting the cooling location away from the \lhto will also reduce boil-off.
In essence, both these actions make the cooling shield\slash tube colder relative to their surroundings (cf.\ Figures~\ref{fig:baseline_temp_A} and~\ref{fig:baseline_temp_B}), meaning that they can intercept more heat.
Another perspective is to consider that the former action reduces the temperature gradient over a fixed thermal path, while the latter action increases the thermal path length between fixed temperatures.

Given the aforementioned trends, it is clear that the maximum theoretical boil-off reduction
for any LAC or BAC system will be achieved by using the lowest possible coolant temperature
and placing the shield\slash tube as far away from the \lhto as possible.\footnote{Of course, using a very cold coolant close to the containment system's outer surface will have practical drawbacks, such as increased power consumption (cf.\ Section~\ref{sec:min_power}).}
Then, for the perlite tank, we observe maximal boil-off reductions of roughly 70\% and 90\% by LAC and BAC, respectively (cf.\ Figures~\ref{fig:boil-off_LAC_A} and~\ref{fig:boil-off_BAC_A}).
For the HePUR-tank, the corresponding reductions are approximately 30\% and 80\% (cf.\ Figures~\ref{fig:boil-off_LAC_B} and~\ref{fig:boil-off_BAC_B}).
The BAC results are in line with the $\sim$80\% heat ingress reduction identified in previous literature~\cite{feller2011doa, xu2020ahl}.
For LAC, directly comparable literature is lacking,
but the results herein appear reconcilable with those by \citet{plachta2014cbr}.

In terms of \emph{relative} heat ingress reduction, both cooling technologies are less effective for the HePUR-tank than for the perlite tank.
In the perlite tank, the primary insulation is so effective that the majority of the heat ingress occurs through the support skirt.
Consequently, LAC applied to the skirt can intercept a significant portion of the heat ingress.
On the other hand, for the HePUR-tank, heat ingress occurs mainly through the primary insulation, making LAC less effective,
as we can observe by comparing Figure~\ref{fig:boil-off_LAC_B} with Figure~\ref{fig:boil-off_LAC_A}.
The distribution of the heat ingress between the skirt and the insulation is less important for BAC,
because the BAC cooling shield encompasses the entire innermost portion of the containment system.
Still, as evidenced by Figures~\ref{fig:boil-off_BAC_A} and~\ref{fig:boil-off_BAC_B},
BAC is still slightly more effective for the perlite tank. 
This is because a fixed thickness of perlite
separating the cooling shield and the inner tank wall  yields larger thermal resistance
than the same thickness of HePUR foam.

In contrast to the above,
it is worth noting that the \emph{absolute} heat ingress reduction is, for the most part, larger for the HePUR-tank.
Indeed, at low $T_{\mathrm{cool}}$ the boil-off reduction by LAC for the HePUR tank (0.07\%/day, cf.\ Figure~\ref{fig:boil-off_LAC_B}) is larger in absolute terms than even a 100\% reduction for the perlite tank (0.04\%/day, cf.\ Section~\ref{sec:baseline_results}).
For the very highest $T_{\mathrm{cool}}$, the case is not as clear cut, and we also observe that some $T_{\mathrm{cool}}$ offer moderate boil-off reduction for the perlite tank but none for the HePUR-tank.
This can be explained by the observation from Section~\ref{sec:baseline_results} that temperatures close to the secondary tank wall tend to be lower for the HePUR-tank than for the perlite tank.
Of course, cooling is only effective if the coolant temperature is lower than the temperature of the passive baseline configuration~1 at the location where cooling is to be installed.
Hence, given our fixed selection of possible cooling locations, the range of effective coolant temperatures is smaller for the HePUR-tank than for the perlite tank.

Another contrastive observation is that
the benefit of placing the LAC tube on the secondary skirt is much larger for the HePUR-tank than for the perlite tank (compare the red lines to the other coloured lines in Figures~\ref{fig:boil-off_LAC_A} and~\ref{fig:boil-off_LAC_B}).
This is because cooling applied to the secondary skirt cools down the secondary tank wall to a much greater extent than when cooling is applied to the primary skirt.
For the perlite tank, where most of the heat enters through the skirt, this secondary effect is relatively minor.
However, for the HePUR tank, it has noticeable impact, since the heat ingress is more evenly distributed.

Finally, we note that, even though the effectiveness of both LAC and BAC
tapers off significantly as $T_{\mathrm{cool}}$ increases,
mid-to-high $T_{\mathrm{cool}}$ can still enable boil-off reduction of practical interest.
For example, $T_{\mathrm{cool}} = \SI{225}{\kelvin}$, which is achievable using 
CO$_2$ as a coolant, 
yields boil-off rate reductions of roughly 25\% and 30\% for the perlite tank with LAC and BAC, respectively.
For the HePUR tank, the relative numbers are less impressive, but in absolute terms,
installation of LAC or BAC with e.g.\ $T_{\mathrm{cool}} = \SI{200}{\kelvin}$
can reduce the boil-off rate by around 0.02\%/day and 0.05\%/day, respectively.

\FloatBarrier

\subsection{Minimizing Power Consumption}
\label{sec:min_power}

In this section, we estimate the minimum power consumption required to avoid net excess boil-off from the perlite-insulated tank and the HePUR-insulated tank.
By ``avoid net excess boil-off'', we mean that the net hydrogen mass flow out of the tank
should exactly match the fuel consumption of any hydrogen-powered systems fuelled from the tank, such as a ship propulsion system.
In this respect, we consider two scenarios: one where no hydrogen is needed for fuel,
such that the target is zero net boil-off,
and one where the target is to reduce boil-off by 50\%,
using the passive baseline configuration~1 as reference.
To this end, we consider LAC and BAC in combination with reliquefaction as needed to meet the target boil-off.
This corresponds to configurations~5 and~6 in Table~\ref{tab:configs}.
Additionally, pure reliquefaction (configurations~3) is included as a baseline.
We consider the same shield\slash tube positions and coolant temperatures as in Section~\ref{sec:min_boil-off}.

\subsubsection{Operating Scenario I: Zero net boil-off}
\label{sec:zero_BOR}

\begin{figure}
  \centering
  \includegraphics[width=0.45\textwidth]{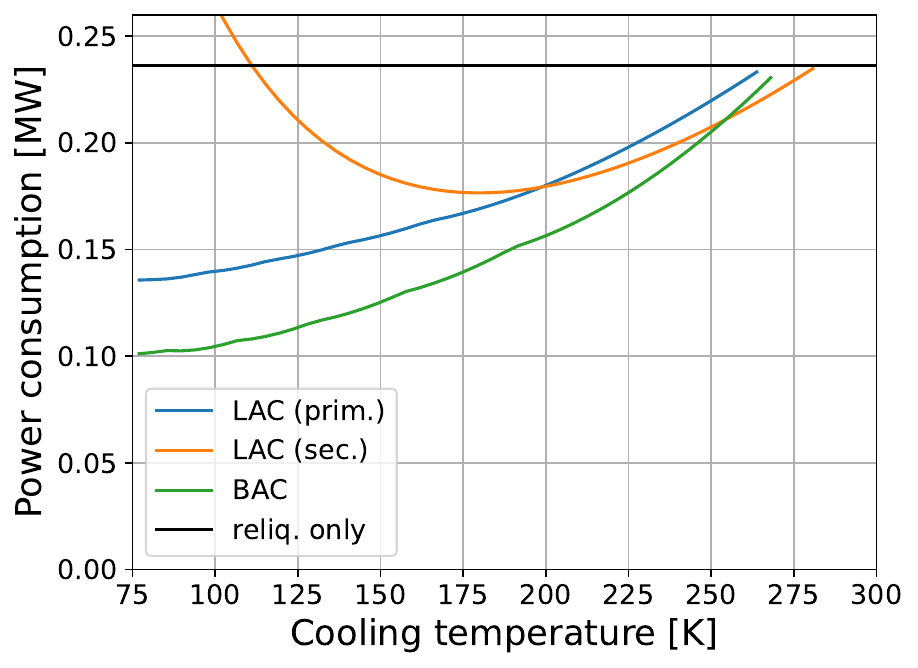}
  \caption{Power consumption required to achieve zero net boil-off for the perlite tank with configurations~4--6, reliquefaction only (black), LAC + reliquefaction (blue/orange: LAC applied to primary/secondary skirt), and BAC + reliquefaction(green).
  }
  \label{fig:power_A_I}
\end{figure}

The power consumption required to achieve \emph{zero net boil-off} for perlite-insulated tank with configurations~4--6 (reliquefaction only, LAC + reliquefaction, and BAC + reliquefaction),
is shown in Figure~\ref{fig:power_A_I} as a function of $T_{\mathrm{cool}}$.\footnote{We emphasize that zero \emph{net} boil-off is not the same as zero boil-off.
Here, any mass flow of evaporated hydrogen out of the tank is compensated by an equal mass flow of reliquefied hydrogen back into the tank, such that the net mass flow is zero.}
For the configurations including LAC installed on the primary skirt (blue) or BAC (green), we display only the minimum power consumption for each $T_{\mathrm{cool}}$,
where the minimum is taken over the seven cooling shield\slash tube positions within the primary insulation space (cf.\ Section~\ref{sec:min_boil-off}).
We consider only a single LAC position on the secondary skirt, and the power consumption for this setup is shown in orange.


First,
we observe that active cooling combined with reliquefaction (represented by the colored lines), by and large,
requires less power than pure reliquefaction (represented by the black line).
In other words, the power consumed by the cooling system is generally 
less than the power saved due to the lower demand for reliquefaction.\footnote{In order to highlight this trade-off, we visualize the cooling power and the reliquefaction power separately for select cases in \ref{app:power_distribution}.}
The only exception to this within the range of $T_{\mathrm{cool}}$ considered here,
is configuration~5 (LAC + reliquefaction) when $T_{\mathrm{cool}}$ is low
and the cooling tube is placed on the secondary skirt (cf.\ the orange line in Figure~\ref{fig:power_A_I}).
In this case, we observe that there exists a minimum power consumption at $T_{\mathrm{cool}} \approx \SI{180}{\kelvin}$.
For $T_{\mathrm{cool}}$ below this temperature,
too much power is spent on cooling that gets lost to the ambient through the secondary skirt and the secondary insulation.
In principle, one should expect such minima for the other cooling setups as well,
but this cannot be observed in Figure~\ref{fig:power_A_I}.
The simple reason for this is that the other minima occur at temperatures below the range considered here.
In \ref{app:extended_results}, we present results for an extended range of $T_{\mathrm{cool}}$,
where the minima are clearly visible for all cases.
Additionally, as demonstrated in \ref{app:efficiency_degradation}, assuming that the cooling system's Carnot efficiency decreases at low temperatures
would push the minima towards higher $T_{\mathrm{cool}}$.

At its ideal temperature of $T_{\mathrm{cool}} = \SI{180}{\kelvin}$,
LAC on the secondary skirt reduces the total power consumption by roughly 25\%
from \SI{0.24}{\mega\watt} to \SI{0.18}{\mega\watt}. At this temperature,
placing the LAC on the primary skirt is only marginally more efficient, while BAC reduces
the total power consumption to \SI{0.14}{\mega\watt}.
The power consumption of the latter two decreases as $T_{\mathrm{cool}}$
is lowered all the way to \SI{77}{\kelvin},
where LAC on the primary skirt yields $P_{\mathrm{tot}} \approx \SI{0.14}{\mega\watt}$,
and BAC yields $P_{\mathrm{tot}} \approx \SI{0.10}{\mega\watt}$.

For higher $T_{\mathrm{cool}}$, the ranking of the setups changes,
with LAC on the secondary skirt becoming increasingly competitive.
Indeed, for $T_{\mathrm{cool}}$ above \SI{255}{\kelvin}, it is actually the most power efficient setup.
Moreover, comparing with the LAC placed on the primary skirt,
the placement on the secondary skirt is superior for $T_{\mathrm{cool}} \gtrsim \SI{200}{\kelvin}$.

\begin{figure}
  \centering
  \includegraphics[width=0.45\textwidth]{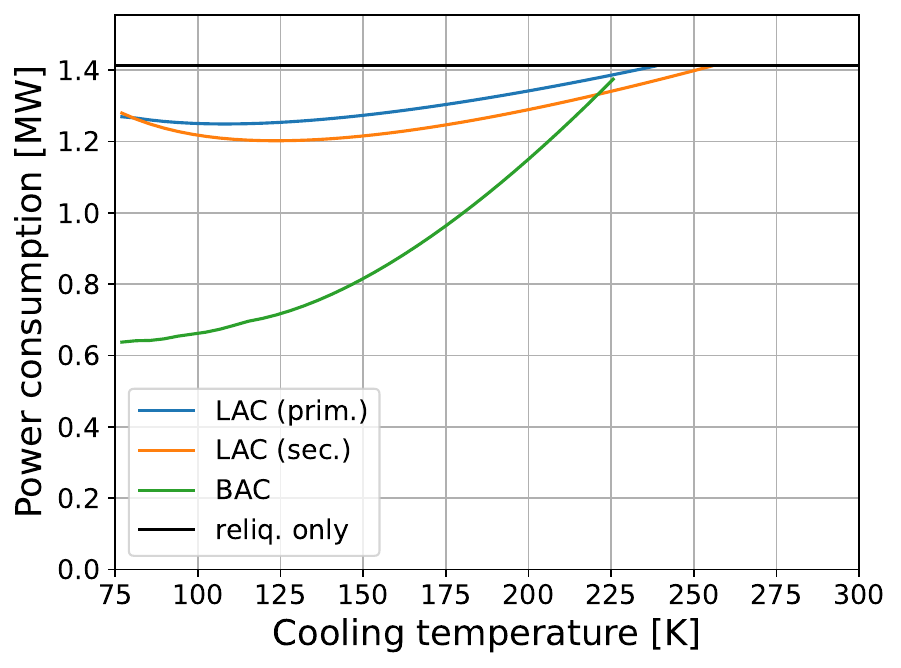}
  \caption{Power consumption required to achieve zero net boil-off for the HePUR tank with configurations~4--6, (reliquefaction only, LAC + reliquefaction, and BAC + reliquefaction).
  The coloring scheme is the same as in Figure~\ref{fig:power_A_I}.}
  \label{fig:power_B_I}
\end{figure}

Figure~\ref{fig:power_B_I} is analogous to Figure~\ref{fig:power_A_I} and displays the total power consumption for the HePUR-insulated tank.
As for the perlite tank, we observe that the inclusion of LAC or BAC is beneficial in comparison to pure reliquefaction.
However, it is also immediately evident that the benefit of BAC over LAC is significantly larger for the HePUR tank.
This is in line with the observations made in Section~\ref{sec:min_boil-off} and is due to the heat ingress being more evenly distributed for the HePUR tank than the perlite tank.
Furthermore, as also noted in Section~\ref{sec:min_boil-off}, this does not imply that LAC should be discounted for the HePUR tank.
In absolute terms, LAC can reduce the power consumption by approximately \SI{0.2}{\mega\watt} (cf.\ the minimum of the orange line at $T_{\mathrm{cool}} \approx \SI{125}{\kelvin}$),
which is more than the largest saving achieved for the perlite tank.



\begin{figure}
  \centering
  \begin{subfigure}{0.23\textwidth}
    \includegraphics[width=\textwidth]{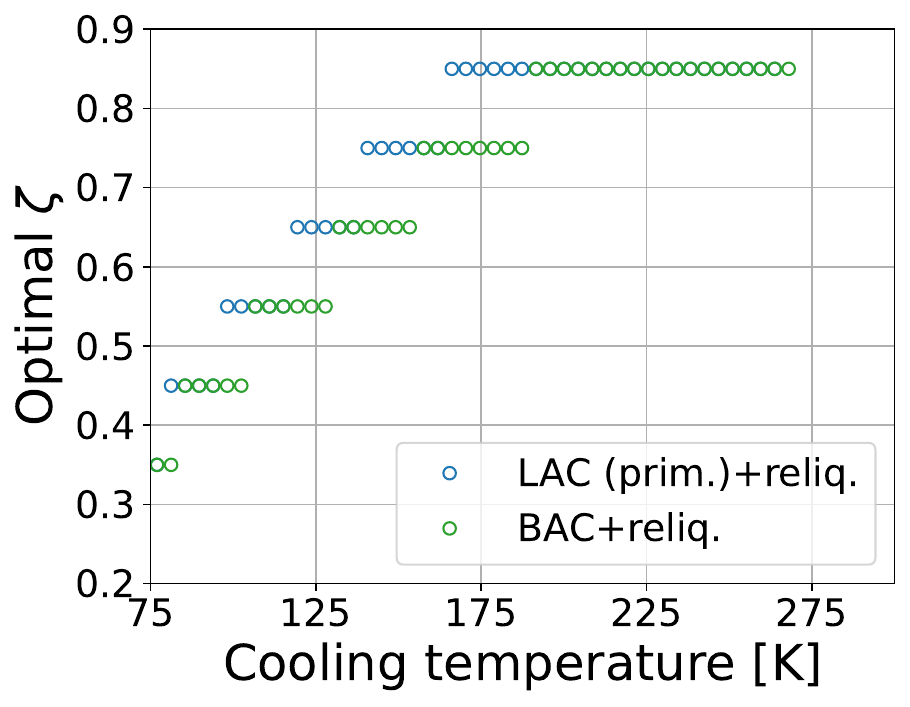}
    \caption{Perlite tank}
    \label{fig:position_A_I}
  \end{subfigure}
  \begin{subfigure}{0.23\textwidth}
    \includegraphics[width=\textwidth]{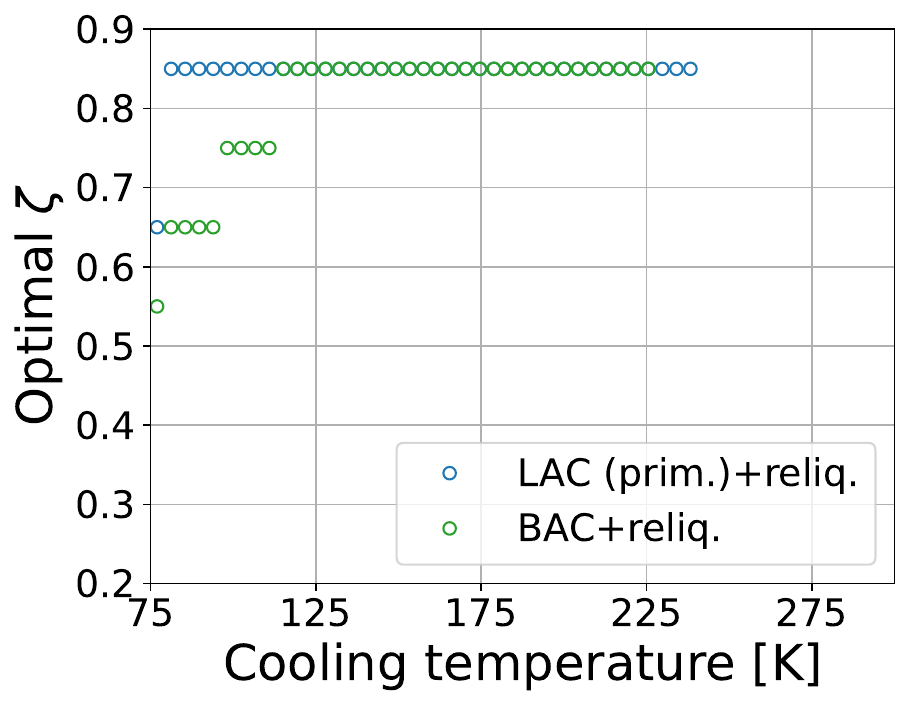}
    \caption{HePUR tank}
    \label{fig:position_B_I}
  \end{subfigure}
  \caption{Optimal cooling shield\slash tube position for achieving zero net boil-off.
  The coloring scheme is the same as in Figure~\ref{fig:power_A_I}.
  Higher $\zeta$-values correspond to placements further away from the contained \lhto.}
  \label{fig:position_I}
\end{figure}

Looking closely at the blue and green curves in Figures~\ref{fig:power_A_I} and~\ref{fig:power_B_I},
one can observe that the curves are not entirely smooth.
The loss of smoothness occurs when the optimal cooling shield\slash tube placement changes.
The optimal placements are shown in Figure~\ref{fig:position_A_I} for the perlite tank, and in Figure~\ref{fig:position_B_I} for the HePUR tank.
A few general trends, which are all consistent with the discussions above, can be observed in these figures:
First, for mid-to-high $T_{\mathrm{cool}}$, it is generally optimal to place the cooling shield\slash tube in the location furthest away from the inner tank wall.
Secondly, as $T_{\mathrm{cool}}$ decreases, the optimal placement shifts towards the inner tank wall in order to reduce the amount of cooling that leaks to the ambient.
Finally, the optimal cooling location for the HePUR tank is generally closer to the inner tank wall than for the perlite tank, other things being equal.
As before, this is because the HePUR tank generally has lower temperatures in the primary insulation space for the passive baseline configuration~1 (cf. Figures~\ref{fig:baseline_temp_A} and~\ref{fig:baseline_temp_B}).


\FloatBarrier

\subsubsection{Operating Scenario II: 50\% Boil-Off Reduction}
\label{sec:halved_BOR}

\begin{figure}
  \centering
  \includegraphics[width=0.45\textwidth]{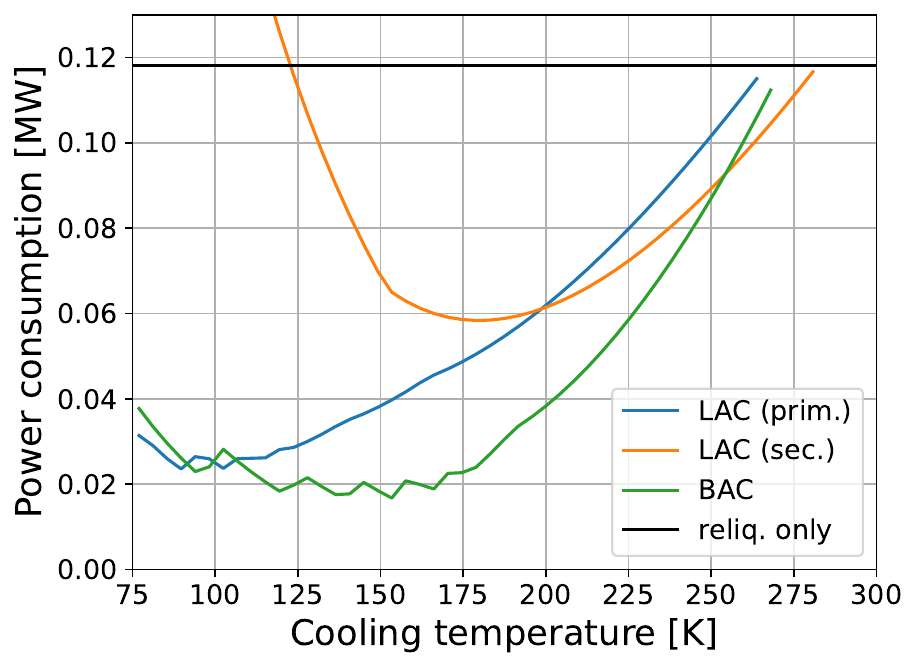}
  \caption{Power consumption required to reduce net boil-off by 50\% for the perlie tank with configurations~4--6 (reliquefaction only, LAC + reliquefaction, and BAC + reliquefaction).
  The coloring scheme is the same as in Figure~\ref{fig:power_A_I}.}
  \label{fig:power_A_II}
\end{figure}

\begin{figure}
  \centering
  \includegraphics[width=0.45\textwidth]{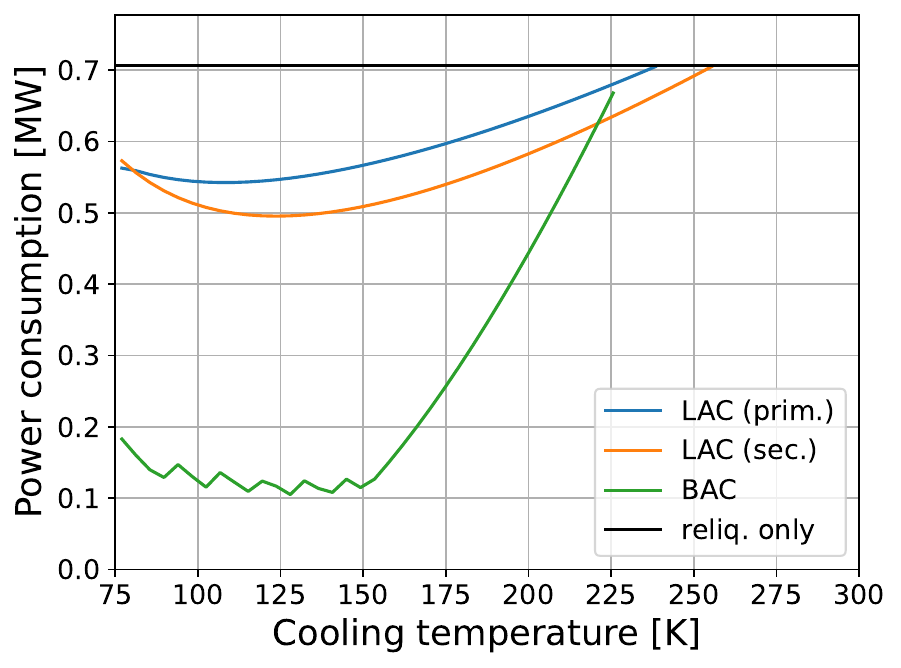}
  \caption{Power consumption required to reduce net boil-off by 50\% for the HePUR tank with configurations~4--6 (reliquefaction only, LAC + reliquefaction, and BAC + reliquefaction).
  The coloring scheme is the same as in Figure~\ref{fig:power_A_I}.}
  \label{fig:power_B_II}
\end{figure}

Figures~\ref{fig:power_A_II} and~\ref{fig:power_B_II} show the power consumption required to \emph{reduce net boil-off by 50\%} for the perlite tank and the HePUR tank, respectively.
In other words, we assume here that hydrogen-powered systems fuelled from the tank require a BOG mass flow half that of the passive baseline configuration~1.
For both tanks, we consider configurations~4--6 (reliquefaction only, LAC + reliquefaction, and BAC + reliquefaction).
In the present scenario, we find that LAC or BAC alone can be sufficient to meet the target boil-off rate for sufficiently low $T_{\mathrm{cool}}$.
In such cases $P_{\mathrm{reliq}} = 0$ and the power consumption is entirely due to the cooling system (cf.\ \ref{app:power_distribution} for further details).

Comparing Figures~\ref{fig:power_A_II} and~\ref{fig:power_B_II} to their scenario~I-counterparts, Figures~\ref{fig:power_A_I} and~\ref{fig:power_B_I},
we first note that the power consumption of pure reliquefaction in Scenario~II is half that of Scenario~I.
This is trivially true, since the mass flow that needs to be reliquefied is halved in Scenario~II.
Furthermore, we observe that applying LAC on the secondary skirt (orange lines)
gives the same power savings as in Scenario~I, and that the corresponding
ideal $T_{\mathrm{cool}}$ is also the same
($T_{\mathrm{cool}} \approx \SI{180}{\kelvin}$ and $T_{\mathrm{cool}} \approx \SI{125}{\kelvin}$ for perlite and HePUR tanks, respectively).
Analogous observations can be made also for the HePUR tank with LAC on the primary skirt (blue line in Figure~\ref{fig:power_B_II}).\footnote{We discuss this observation further in \ref{app:power_distribution}.}

For the other three setups, optimal performance is achieved by active cooling alone.
Moreover, especially for BAC (green lines), there exists a large number of $(T_{\mathrm{cool}}, \zeta)$-pairs that offer practically the same total power consumption,
as evidenced by the power consumption plateaus for $T_{\mathrm{cool}} \in [\SI{120}{\kelvin}, \SI{160}{\kelvin}]$ in Figure~\ref{fig:power_A_II} and for $T_{\mathrm{cool}} \in [\SI{100}{\kelvin}, \SI{150}{\kelvin}]$ in Figure~\ref{fig:power_B_II}.
These plateaus are somewhat jagged in the figures, which is because we only consider a relatively small number of shield positions.
Increasing the number of considered shield positions would smooth out the plateaus.
From a design perspective, these plateaus enable a trade-off between shield mass and coolant temperature.

\begin{figure}
  \centering
  \begin{subfigure}{0.23\textwidth}
    \includegraphics[width=\textwidth]{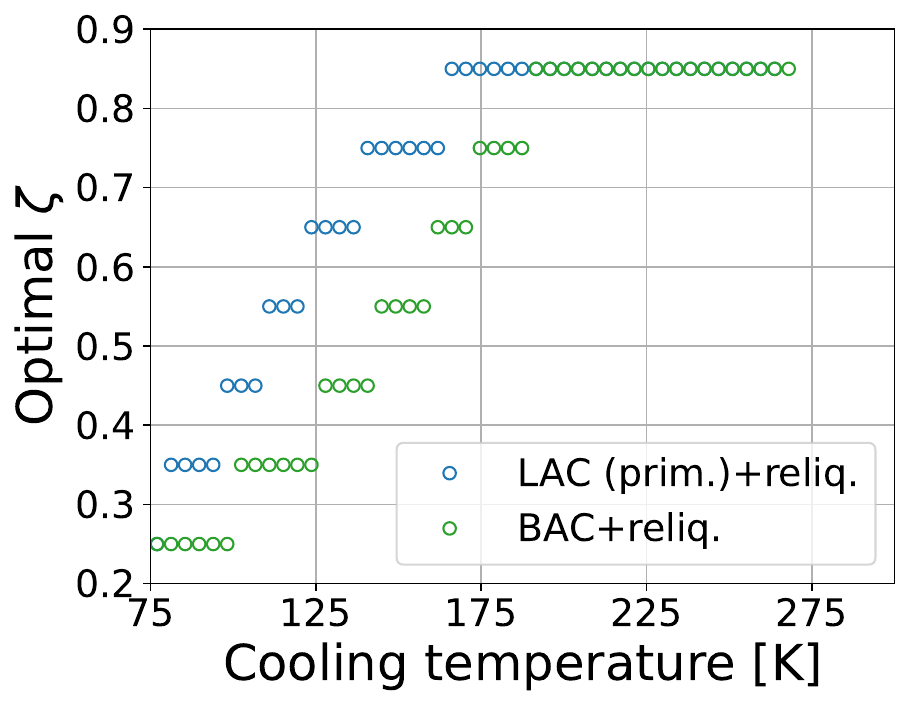}
    \caption{Perlite tank}
    \label{fig:position_A_II}
  \end{subfigure}
  \begin{subfigure}{0.23\textwidth}
    \includegraphics[width=\textwidth]{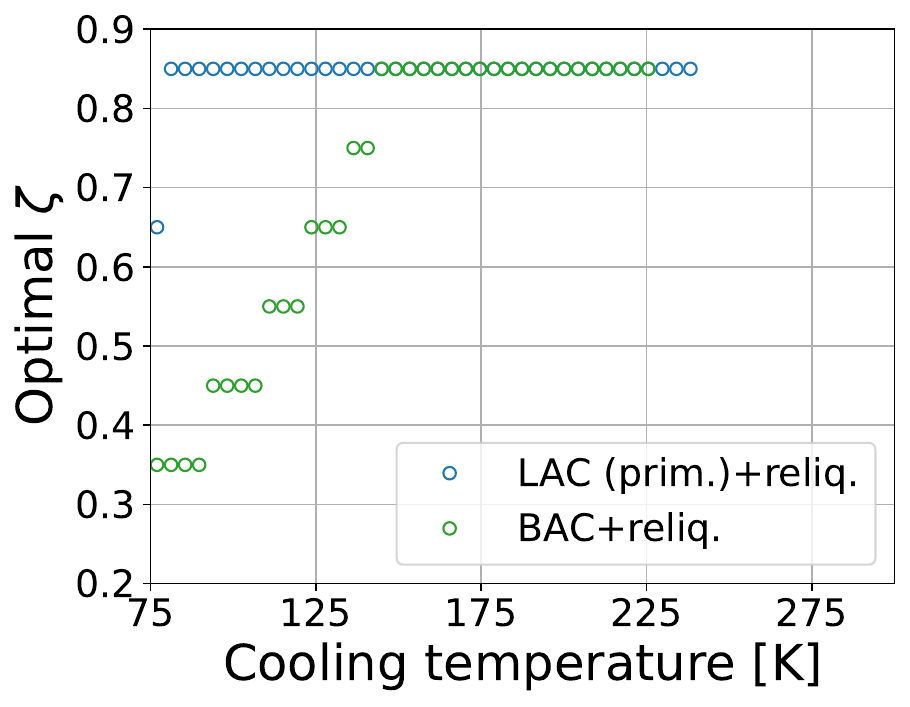}
    \caption{HePUR tank}
    \label{fig:position_B_II}
  \end{subfigure}
  \caption{Optimal cooling shield\slash tube positions for achieving a 50\% reduction in net boil-off.
  The coloring styles are the same as in Figure~\ref{fig:position_I}.}
  \label{fig:position_II}
\end{figure}

Figure~\ref{fig:position_II} shows how the optimal cooling shield\slash tube placement varies with $T_{\mathrm{cool}}$ in Scenario~II.
Comparing with Figure~\ref{fig:position_I} for Scenario~I,
we observe that shield\slash tube positions closer to the inner tank wall are generally more favorable in Scenario~II.
However, this mainly applies to comparatively low $T_{\mathrm{cool}}$.
For such $T_{\mathrm{cool}}$, moving the shield\slash tube closer to the inner tank wall enables sufficient cooling to be maintained while the amount of cooling that leaks to the ambient is reduced.
For higher $T_{\mathrm{cool}}$, we do not observe significant differences between the two scenarios.
Placements away from the inner tank wall are then generally favorable in both scenarios.

\section{Discussion}
\label{sec:discussion}

\subsection{Practical Considerations}
\label{sec:in_practice}

The results presented in Sections~\ref{sec:min_boil-off} and~\ref{sec:min_power}
clearly indicate that BAC with a low coolant temperature has the greatest theoretical potential for reducing both boil-off generation and power consumption.
However, LAC also achieved impressive performance in many cases,
and it is important to bear in mind that theoretical performance is not
the only matter to consider when evaluating active cooling systems.
Here, we discuss other practical matters relevant to active cooling 
of \lhto tanks.


First, one of the most prominent differences between LAC and BAC is the difference in their size.
At the smaller scales for which BAC has most commonly been considered,
the size difference has only minor practical implications.
However, for the tank size considered in the present work, the BAC shield will have an area
of approximately \SI{6000}{\meter^2} and, even at a thickness of only a couple of millimeters,
its weight will be close to 100 metric tonnes.
On the other hand, LAC in essence only requires a single cooling tube running
along the circumference of the support skirt, which is some 130 meters
for the present tank design.
The order of magnitude difference in size favors LAC in terms of manufacturing costs,
coolant management complexity, and installation complexity. 
For the latter, the less significant interference with installation of other components, such as the primary insulation, also plays in favor of LAC.
Similarly, an LAC tube does not interfere with manual tank wall inspections like a BAC shield would.

Another aspect to consider is that both LAC and BAC may cause
increased thermal stress within the tank's structural elements.
This is especially relevant when a low $T_{\mathrm{cool}}$ is applied
close to the secondary tank wall, where the temperature would otherwise be relatively high.
In this case, cold leaking to the ship's inner hull may also be of concern.
In addition to causing thermal stress within the hull structure,
this could conceivably also cause frost problems in severe cases.
For these reasons, in order to preserve structural integrity,
it may be necessary to use a higher $T_{\mathrm{cool}}$ than
is optimal from a purely thermal perspective.
Hence, we recommend that the structral impact of active cooling be
explored in detail in future work.

Also in favor of higher $T_{\mathrm{cool}}$ is the current state-of-the-art
of refrigeration systems. Generally, refrigeration systems operating at a higher
$T_{\mathrm{cool}}$ are more efficient, have greater capacity, and are cheaper.
Increasing $T_{\mathrm{cool}}$ also broadens the available selection of refrigerants.
For example, for $T_{\mathrm{cool}} \gtrsim \SI{225}{\kelvin}$ (liquid) CO$_2$ can be used,
with benefits including non-flammability, affordability, equipment compactness and technological maturity.
Additionally, avoiding heat leaks through coolant transfer lines
becomes easier and cheaper as $T_{\mathrm{cool}}$ increases.
Should lower $T_{\mathrm{cool}}$ still prove desirable,
nitrogen could be an attractive choice of refrigerant for $T_{\mathrm{cool}}$ all the way down to \SI{77}{\kelvin}.

Finally, we highlight some attractive benefits of cooling installed outside the secondary tank wall.
For one, it is then easier to access the cooling system, both for installation and eventual maintenance.
Secondly, coolant leaks are less of a concern outside the secondary tank wall than within,
both because the consequences are less severe and because they are easier to fix.
Coolant leaks inside the primary insulation space would be especially detrimental
for tank designs using evacuated insulations, such as the evacuated perlite considered in the present work.

\subsection{Extensions of the Present Work}
\label{sec:extensions}

Previously, we recommended that the structural implications of specific active cooling
setups be explored in future studies.
In this section, we discuss additional extensions of the
present work that would also generate useful insights into the
applicability and benefits of LAC and BAC for large-scale hydrogen storage tanks.

An important limitation of the present work is that we have only considered
steady-state operation with an approximately full tank.
Especially for transport tanks, such as considered herein,
performance at low filling levels is also highly relevant.
To this end, one could for example prescribe a piece-wise linear temperature profile within the ullage space, as suggested by \citet{wang2024tmo}.
Additionally, it may be prudent to consider performance during loading\slash unloading,
even though the time spent on these operations is generally small compared to the time spent in transit.
Performance during warm-up\slash cool-down in connection with tank inspections could also be of interest.

The analyses in Sections~\ref{sec:min_boil-off} and~\ref{sec:min_power}
consider only cooling setups with a single cooling shield or a single cooling tube.
Installing two or more tubes\slash shields cooled to different temperatures could
concievably yield increased efficiency, as suggested in previous literature~\cite{yu2023dao, kim2000tda} on BAC.
Exploring the potential of such solutions for LAC could be an interesting avenue for future work.

In Scenario~II, and in general when BOG is required as fuel,
exergy can be extracted from that BOG in order to provide improved cooling.
This could be achieved by directly cooling the tank via
one or more vapor-cooled shields or tubes.
Alternatively, the BOG could be fed through a heat exchanger where it cools, and possibly reliquefies,
a secondary fluid (e.g.\ CO$_2$ or N$_2$) acting as the coolant of an active cooling system.
These options are not considered in the present work, and could serve to lower
BOG generation and power consumption beyond the numbers presented herein.

In this work, we have assumed that the temperature inside the tank's holding space
is fixed and hence independent of any active cooling.
Yet, as noted in Section~\ref{sec:results}, we observe that some cooling setups
result in significant cold leak through the secondary skirt and the secondary insulation.
Given our modelling assumptions, the power consumption corresponding to this leak is wasted.
However, so long as the heat removed by the cryocooler is dumped outside the holding space,
the leak will in practice reduce the temperature in the holding space,
which in turn has a noticeable impact on the tank's heat ingress~\cite{aasen2024tpe}.
Consequently, the seemingly wasted power will not be entirely wasted in practice,
meaning that the potential of LAC and BAC to reduce boil-off generation may have
been moderately underestimated in Section~\ref{sec:results}.
Quantitative estimates of this effect would require a detailed
geometric description of the holding space and its interfaces towards the
ship's deck and inner hull. They are therefore beyond the scope of the present work.

Finally, we highlight that boil-off rate and power consumption do
not provide a complete view on the economic benefit of active cooling.
Additional operating expenses, e.g.\ due to maintenance,
and capital expenses from the system design and installation,
have not been included in our quantitative analyses.
However, the qualitative discussion in Section~\ref{sec:in_practice} indicates that
LAC will compare favorably to BAC with respect to these expenses.

\section{Conclusions}
\label{sec:conclusions}

In the present work, we have compared the potential of local area cooling (LAC)
and broad area cooling (BAC) to reduce boil-off generation from liquid hydrogen (\lhto) storage tanks.
Additionally, we compare the minimum power consumption required for LAC and BAC, possibly
supplemented with reliquefaction, to achieve specified net boil-off rates.
The analyses have been conducted in the context of ship-borne \lhto transport
using a spherical, double-walled design for a \SI{40000}{\meter^3}-capacity, skirt-supported \lhto storage tank.
We consider two primary insulation material choices: evacuated perlite
and helium-filled polyurethane (HePUR) foam.
When no active cooling is applied, these insulation materials give
boil-off rates of 0.04\%/day and 0.24\%/day, respectively.

We find that LAC and BAC with liquid nitrogen as coolant
can reduce boil-off by up to 70\% and 90\%, respectively,
for the perlite-insulated tank.
The corresponding numbers are 25\% and 30\% when the coolant is liquid CO$_2$.
For the less performant HePUR insulation, cooling with liquid nitrogen
can give a boil-off reduction of 30\% for LAC and 80\% for BAC,
while CO$_2$ cooling gives less than 10\% reduction for both LAC and BAC.
Even though the relative savings are less impressive for the HePUR-insulated tank, especially for LAC,
the absolute savings (0.02--0.07\%/day) are still economically significant.

As for power power consumption, our analysis suggests that
LAC can reduce the power needed to achieve zero net boil-off
by up to \SI{0.1}{\mega\watt} for the perlite-insulated tank
and up to \SI{0.2}{\mega\watt} for the HePUR-insulated tank
when comparing to reliquefaction only.
The analogous numbers for BAC are \SI{0.14}{\mega\watt} and \SI{0.75}{\mega\watt}.
Moreover, a 50\% reduction in net boil-off from the perlite-insulated tank requires \SI{0.1}{\mega\watt} less power
when either LAC or BAC is applied.
For the HePUR-insulated tank, the same boil-off reduction becomes \SI{0.2}{\mega\watt} cheaper with LAC
and \SI{0.6}{\mega\watt} cheaper with BAC.

Overall, our results indicate that both LAC and BAC enable
practically significant boil-off and power consumption savings.
In terms of thermal performance, BAC has an edge over LAC.
However, LAC has several practical benefits that can make it a competetive option,
especially in early-generation transport ships.
These benefits include smaller size and comparatively simple installation, inspection, maintenance and coolant management.
From an operational point of view, LAC installed on the outside of the secondary tank wall
is especially promising for tanks with evacuated primary insulation.

\section*{Acknowledgments}

This publication is based on results from the research project \lhto
Pioneer -- Ultra-insulated seaborne containment system for global \lhto
ship transport, performed under the ENERGIX programme. The authors
acknowledge the following parties for financial support: Gassco,
Equinor, Air Liquide, HD Korea Shipbuilding \& Offshore Engineering,
Moss Maritime and the Research Council of Norway (320233).
Additionally, the authors extend their gratitude to
Dr.\ Magnus Aa. Gjennestad for valuable discussions and input.

\section*{CRediT authorship contribution statement}

\textbf{Sindre Stenen Blakseth:} Conceptualization, Methodology, Software, Formal Analysis, Investigation, Writing -- Original Draft, Visualization;
\textbf{Ailo Aasen:} Conceptualization, Methodology, Formal Analysis, Data Curation, Writing -- Review \& Editing, Supervision;
\textbf{André Massing:} Methodology, Writing -- Review \& Editing, Supervision, Funding Acquisition;
\textbf{Petter Nekså:} Conceptualization, Methodology, Formal Analysis, Writing -- Review \& Editing, Supervision, Funding Acquisition.


\appendix

\section{Grid Refinement Study}
\label{app:grid_refinement}

A grid refinement study has been conducted in order to identify an adequate
resolution for the computational mesh used by the finite element method described in Section~\ref{sec:modelling_temperatures}.
The study was conducted by computing the heat ingress $Q_{\mathrm{in}}$ of the passive
baseline configuration~1 using increasingly high mesh resolutions.
The progressive mesh refinement was achieved using Netgen's in-built
\verb|mesh.Refine()| functionality, which approximately halves
the mesh element size in each spatial dimension every time it is applied.

\begin{figure}
  \centering
  \begin{subfigure}{0.45\textwidth}
    \includegraphics[width=\textwidth]{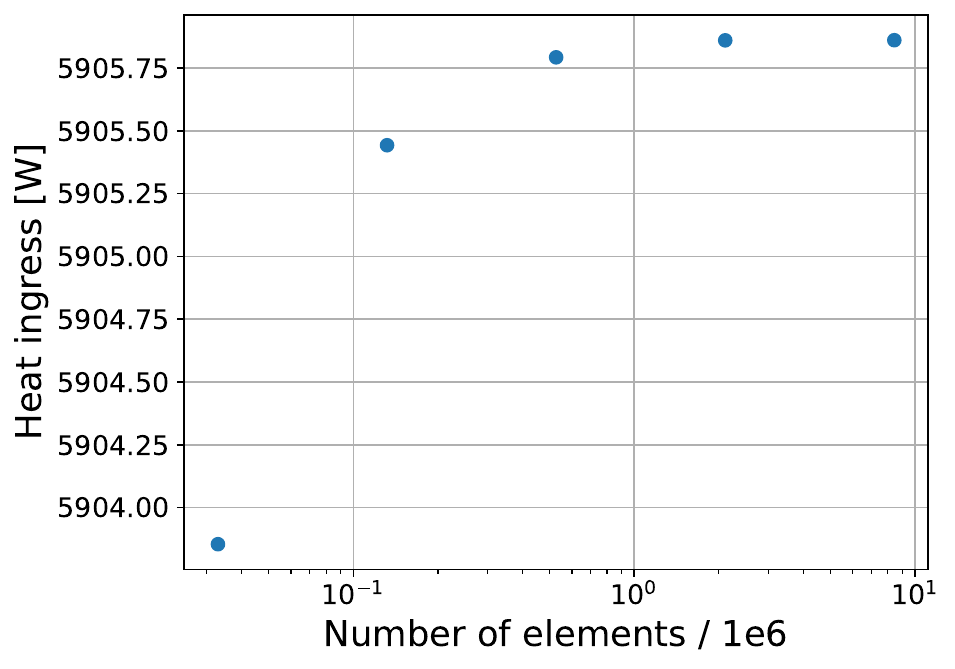}
    \vspace{-1.5em}
    \caption{Perlite tank}
    \label{fig:baseline_conv_A}
  \end{subfigure}\\
  \vspace{1.0em}
  \begin{subfigure}{0.45\textwidth}
    \includegraphics[width=\textwidth]{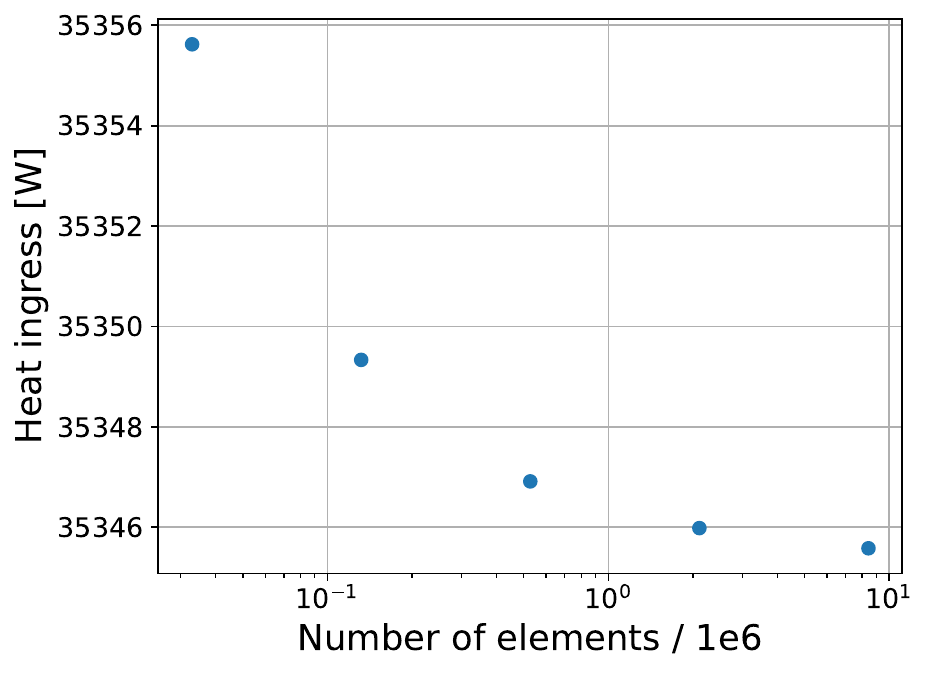}
    \vspace{-1.5em}
    \caption{HePUR tank}
    \label{fig:baseline_conv_B}
  \end{subfigure}
  \caption{The heat ingress $Q_{\mathrm{in}}$ of the passively cooled baseline containment systems, as computed on increasingly fine finite element meshes.}
  \label{fig:baseline_conv}
\end{figure}

Figure~\ref{fig:baseline_conv} shows, for both the perlite tank and the HePUR tank, the estimated $Q_{\mathrm{in}}$
as a function of the number of (first-order, triangular) elements in the mesh.
We observe that the estimates change relatively little from one mesh resolution to the next.
The second-coarsest mesh already enables $Q_{\mathrm{in}}$ estimates
that are within 0.01\% of the corresponding estimates
obtained using the finest mesh.
From this observation, we conclude that the second-coarsest mesh
is sufficient to produce practically meaningful and trustworthy results.
At the same time, this mesh contains few enough elements that the
$2\cdot(7+7+1)\cdot30 = 900$ model evaluations required to produce the results of Section~\ref{sec:results}
were not exceedingly costly in terms of computational power.

\FloatBarrier

\section{Distribution of Power Consumption}
\label{app:power_distribution}

Configurations~5 or~6 in Table~\ref{tab:configs} (LAC/BAC + reliquefaction) each has a total power consumption given by
$P_{\mathrm{tot}} = P_{\mathrm{cool}} + P_{\mathrm{reliq}}$,
where $P_{\mathrm{cool}}$ is the power consumption of the LAC or BAC cooling system,
and $P_{\mathrm{reliq}}$ is the power required to reliquefy any excess boil-off.
The dependency of $P_{\mathrm{tot}}$ on the coolant temperature $T_{\mathrm{cool}}$
has been studied in detail in Section~\ref{sec:min_power}.
Here, we complement the previously presented results by visualizing the two components of $P_{\mathrm{tot}}$ separately for select cases.

\begin{figure}
  \centering
  \includegraphics[width=0.45\textwidth]{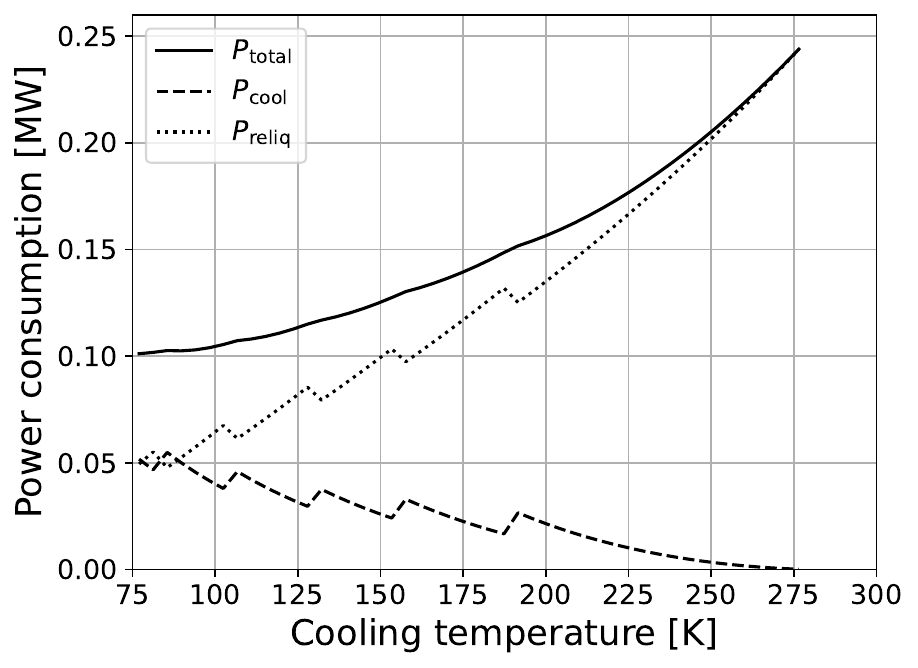}
  \caption{Contributions to the power consumption for the perlite tank with Configuration~6 (BAC and reliquefaction) in Scenario~I.
  The total power consumption (solid line) is the same as the solid green line in Figure~\ref{fig:power_A_I}.}
  \label{fig:power_distribution_A_I_BAC}
\end{figure}

\begin{figure}
  \centering
  \includegraphics[width=0.45\textwidth]{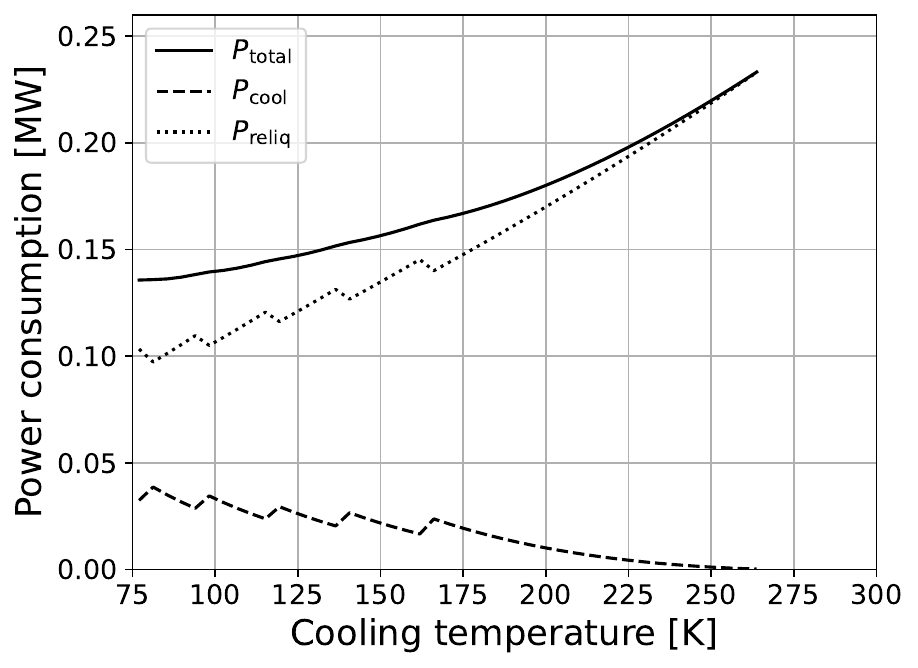}
  \caption{Contributions to the power consumption for the perlite tank with Configuration~5 (LAC and reliquefaction) in Scenario~I.
  The cooling tube is placed on the primary skirt,
  meaning that the total power consumption (solid line) is the same as the solid blue line in Figure~\ref{fig:power_A_I}.}
  \label{fig:power_distribution_A_I_LAC_primary}
\end{figure}

\begin{figure}
  \centering
  \includegraphics[width=0.45\textwidth]{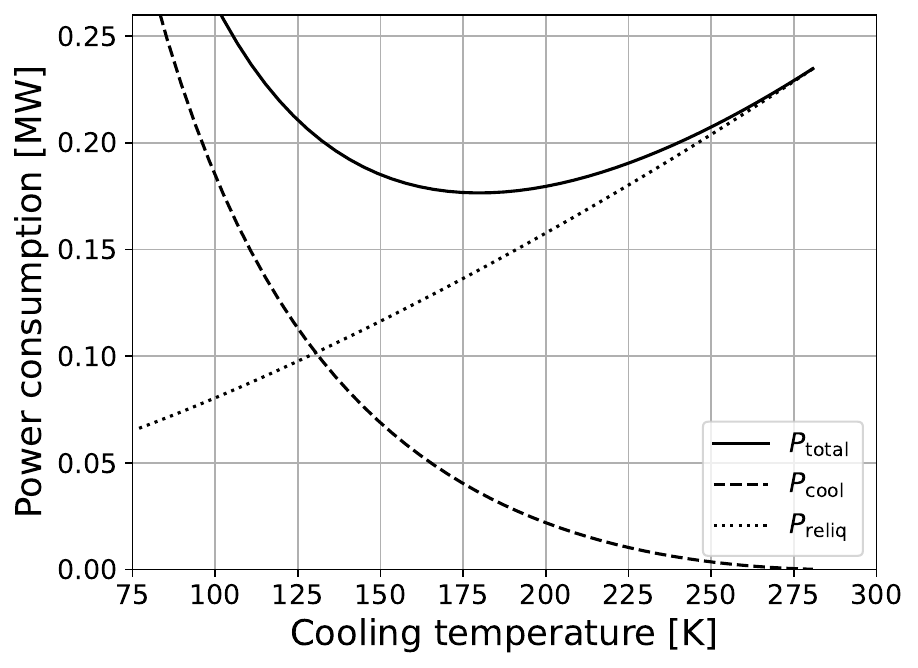}
  \caption{Contributions to the power consumption for the perlite tank with Configuration~5 (LAC and reliquefaction) in Scenario~I.
  The cooling tube is placed on the secondary skirt,
  meaning that the total power consumption (solid line) is the same as the solid orange line in Figure~\ref{fig:power_A_I}.}
  \label{fig:power_distribution_A_I_LAC_secondary}
\end{figure}

Figures~\ref{fig:power_distribution_A_I_BAC}--\ref{fig:power_distribution_A_I_LAC_secondary}
extend the Scenario~I-results presented for the perlite tank with configurations~5 and~6 in Figure~\ref{fig:power_A_I}.
For BAC (Figure~\ref{fig:power_distribution_A_I_BAC}) and LAC applied to the \emph{primary} skirt (Figure~\ref{fig:power_distribution_A_I_LAC_primary}),
we observe that $P_{\mathrm{cool}}$ and $P_{\mathrm{reliq}}$ follow similar trends.
In both cases, $P_{\mathrm{reliq}}$ is generally larger than $P_{\mathrm{cool}}$
(though the difference between them is greater for the LAC than the BAC),
and the two become gradually more equal as $T_{\mathrm{cool}}$ is reduced.
Moreover, $P_{\mathrm{cool}}$ and $P_{\mathrm{reliq}}$ are noticeably jagged functions
of $T_{\mathrm{cool}}$, and the jumps correspond to shifts in optimal shield\slash tube placements (cf.\ Figure~\ref{fig:position_I}).

For the perlite tank with LAC on the secondary skirt, we observe from Figure~\ref{fig:power_distribution_A_I_LAC_secondary}
that $P_{\mathrm{cool}}$ increases drastically for low $T_{\mathrm{cool}}$.
This is caused by significant cold leakage through the secondary skirt and the secondary insulation.
As discussed in Section~\ref{sec:min_power}, this results in
a minimal power consumption at $T_{\mathrm{cool}} \approx \SI{180}{\kelvin}$
for this setup.

\begin{figure}
  \centering
  \includegraphics[width=0.45\textwidth]{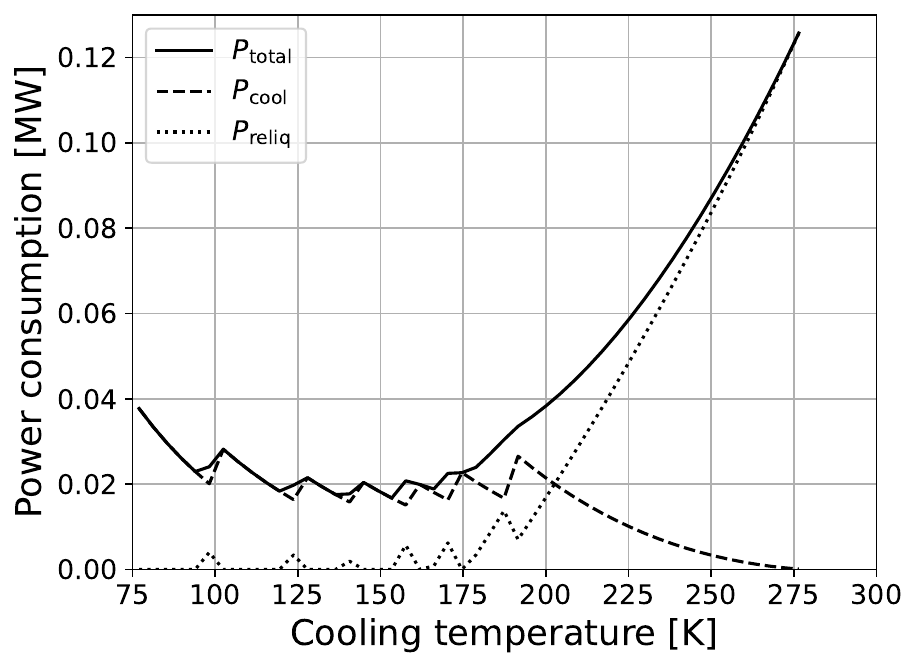}
  \caption{Contributions to the power consumption for the perlite with Configuration~6 (BAC and reliquefaction) in Scenario~II.
  The total power consumption (solid line) is the same as the solid green line in Figure~\ref{fig:power_A_II}.}
  \label{fig:power_distribution_A_II_BAC}
\end{figure}

\begin{figure}
  \centering
  \includegraphics[width=0.45\textwidth]{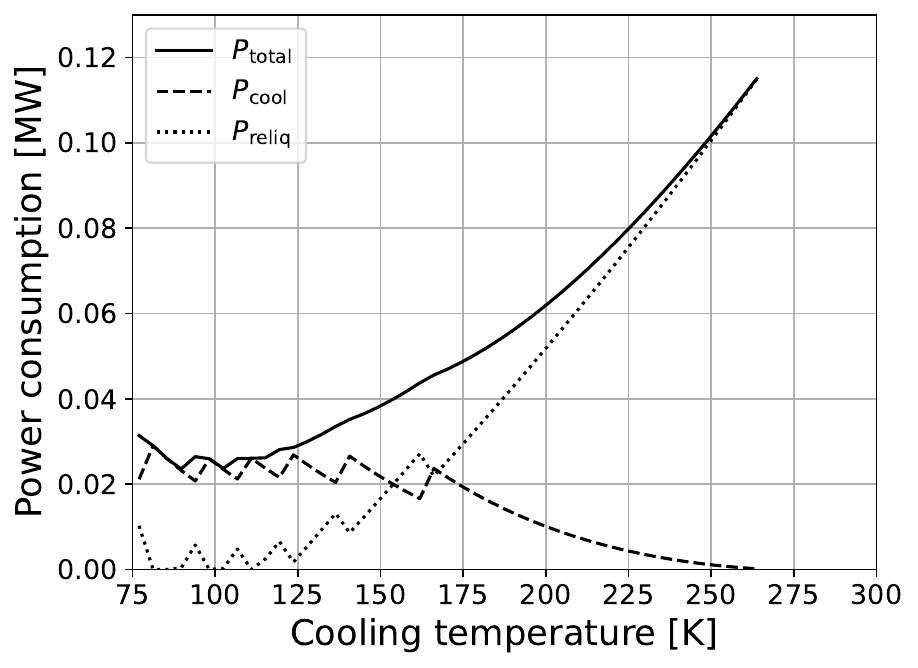}
  \caption{Contributions to the power consumption for the perlite tank with Configuration~5 (LAC and reliquefaction) in Scenario~II.
  The cooling tube is placed on the primary skirt,
  meaning that the total power consumption (solid line) is the same as the solid blue line in Figure~\ref{fig:power_A_II}.}
  \label{fig:power_distribution_A_II_LAC_primary}
\end{figure}

\begin{figure}
  \centering
  \includegraphics[width=0.45\textwidth]{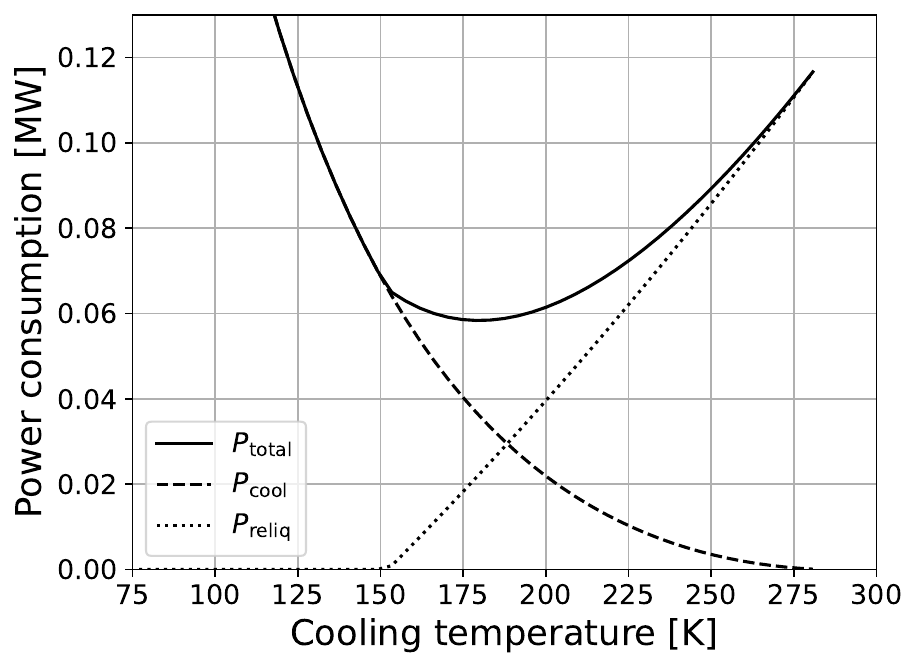}
  \caption{Contributions to the power consumption for the perlite tank with Configuration~5 (LAC and reliquefaction) in Scenario~II.
  The cooling tube is placed on the secondary skirt,
  meaning that the total power consumption (solid line) is the same as the solid orange line in Figure~\ref{fig:power_A_II}.}
  \label{fig:power_distribution_A_II_LAC_secondary}
\end{figure}

The Scenario~II-results originally presented in Figure~\ref{fig:power_A_II}
are extended in Figures~\ref{fig:power_distribution_A_II_BAC}--\ref{fig:power_distribution_A_II_LAC_secondary}.
The most notable feature here is that $P_{\mathrm{reliq}}$
goes to zero in all three figures.
This implies that, for sufficiently low $T_{\mathrm{cool}}$, the active cooling alone is sufficient to achieve the target 50\% reduction in net boil-off.\footnote{In Figures~\ref{fig:power_distribution_A_II_BAC} and~\ref{fig:power_distribution_A_II_LAC_primary},
$P_{\mathrm{cool}}$ oscillates between being zero and non-zero over a range of $T_{\mathrm{cool}}$-values.
This is a numerical artifact resulting from the low number of shield\slash tube positions considered in the present work.}

Now, let us consider why LAC on the primary skirt for the HePUR tank, and LAC on the secondary skirt for both tank,
yield the same optimal $T_{\mathrm{cool}}$ and the same maximal power reduction in Scenario~II as in Scenario~I (cf.\ Section~\ref{sec:halved_BOR}).
To this end, we introduce $T^*$ as the highest $T_{\mathrm{cool}}$ that makes active cooling alone sufficient to meet the target boil-off rate.\footnote{If no such $T_{\mathrm{cool}}$ exists, setting $T^* = 0$ works well for the sake of the argument.}
For example, according to Figure~\ref{fig:power_distribution_A_II_LAC_secondary}, $T^* \approx \SI{150}{\kelvin}$ for the perlite tank with LAC on the secondary skirt.
For $T_{\mathrm{cool}} > T^*$, there will be a trade-off between the power consumption of the cooling system and the power savings due to reduced reliquefaction.
This trade-off is the same as in Scenario~I, meaning that the total power \emph{saving} for any particular setup with $T_{\mathrm{cool}} > T^*$ is then the same in both scenarios.

What the three setups mentioned above have in common is that $T^*$ is lower than the optimal $T_{\mathrm{cool}}$ from Scenario~I.
As a consequence, the minima identified in Scenario~I are still present in Scenario~II for these setups.
For LAC on the secondary skirt, we only consider a single position, so there is no way that lowering $T_{\mathrm{cool}}$ below $T^*$ can reduce power consumption.
For LAC on the primary skirt, moving the cooling tube towards the inner tank could conceivably result in a second local minimum at a lower $T_{\mathrm{cool}}$,
but this happens not to be the case in Figure~\ref{fig:power_B_II}.
Hence, both the optimal $T_{\mathrm{cool}}$ and the maximum power saving do not change from Scenario~I to Scenario~II for the aforementioned three setups.
For the remaining three setups (the perlite tank with BAC and LAC on the primary skirt, and the HePUR tank with BAC),
$T^*$ is higher than the optimal $T_{\mathrm{cool}}$ from Scenario~I.
Then, optimal performance is achieved by LAC or BAC alone.

\FloatBarrier

\section{Extended Power Consumption Results}
\label{app:extended_results}

In this section, we extend the power consumption results from Section~\ref{sec:min_power} in two ways.
Firstly, we expand the range of $T_{\mathrm{cool}}$ from $[\SI{77}{\kelvin}, \SI{285}{\kelvin}]$ to $[\SI{22}{\kelvin}, \SI{285}{\kelvin}]$.
Secondly, we include direct hydrogen cooling (DHC) as an alternative to reliquefaction.
To achieve a target net boil-off rate $\dot{m}_{\mathrm{fuel}}$,
the DHC system must remove the heat
\begin{equation}
  Q_{\mathrm{excess}} = \mathrm{max}(Q_{\mathrm{in}} - \Delta H_{\mathrm{vap}} \dot{m}_{\mathrm{fuel}}, 0)
\end{equation}
from the contained \lhto.
This results in a power consumption given by
\begin{equation}
  P_{\textsc{dhc}} = \frac{Q_{\mathrm{excess}}}{\eta_{\textsc{dhc}}\frac{T_{\mathrm{cool},\textsc{dhc}}}{T_{\mathrm{amb}} - T_{\mathrm{cool},\textsc{dhc}}}},
  \label{eq:power_dhc}
\end{equation}
where $\eta_{\textsc{dhc}}$ is the Carnot efficiency of the DHC system's refrigeration unit and $T_{\mathrm{cool},\textsc{dhc}}$ is the temperature at which cooling is supplied.
Distribution of the cooling from the refrigeration unit to the contained \lhto can be realized e.g.\
by extracting \lhto from the tank, which is then sub-cooled and returned to the tank.
Then, $P_{\textsc{dhc}}$ can be estimated using $\eta_{\textsc{dhc}} = 0.2$~\cite{deserranno2014ooa, notardonato2017zbm} and $T_{\mathrm{cool}, \textsc{dhc}} = \SI{18}{\kelvin}$.

\begin{figure}
  \centering
  \includegraphics[width=0.45\textwidth]{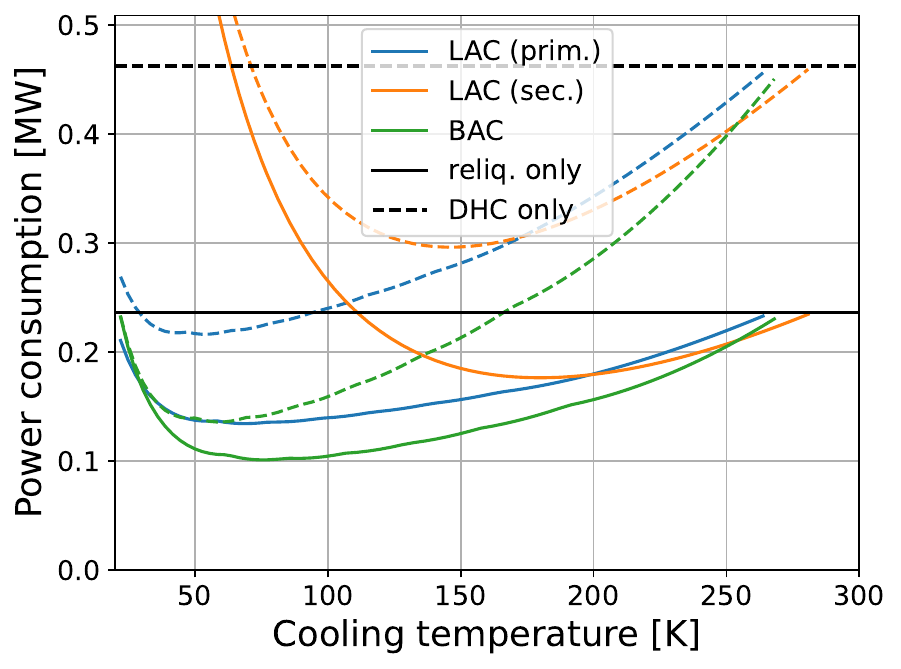}
  \caption{Power consumption required to achieve net zero boil-off for the perlite tank.
  This figure is the same as Figure~\ref{fig:power_A_I}, except that the $x$-axis has here been extended to included lower temperatures.}
  \label{fig:power_A_I_eta_const}
\end{figure}

\begin{figure}
  \centering
  \includegraphics[width=0.45\textwidth]{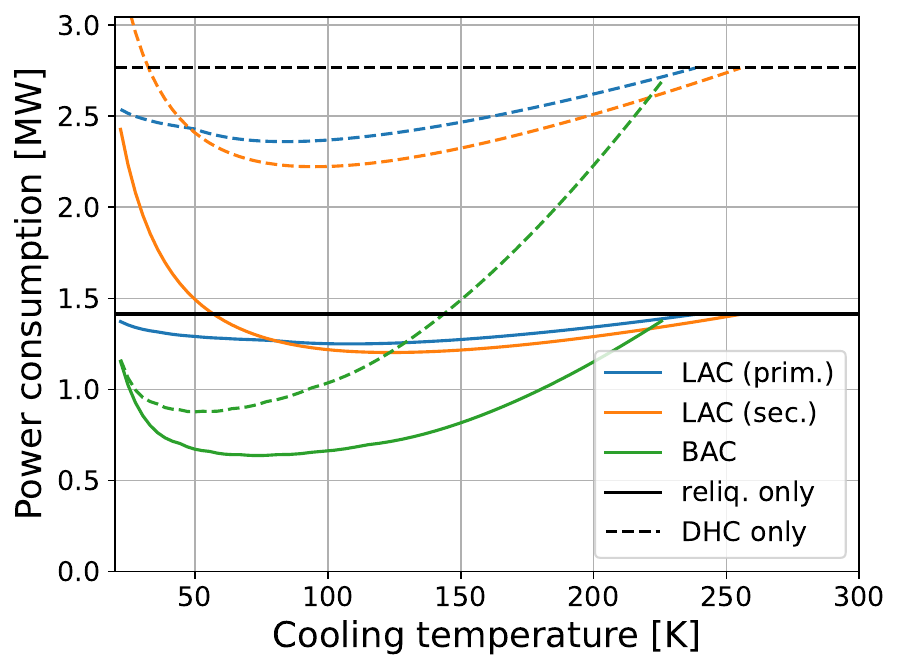}
  \caption{Power consumption required to achieve net zero boil-off for the HePUR tank.
  This figure is the same as Figure~\ref{fig:power_B_I}, except that the $x$-axis has here been extended to included lower temperatures,
  and results for cooling configurations with DHC have been included.}
  \label{fig:power_B_I_eta_const}
\end{figure}

\begin{figure}
  \centering
  \includegraphics[width=0.45\textwidth]{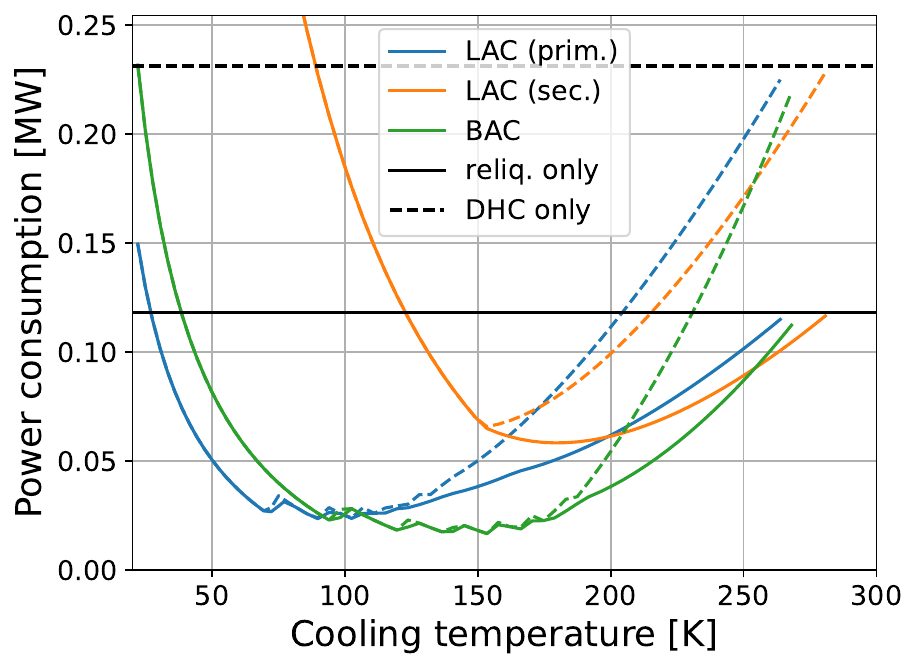}
  \caption{Power consumption required to achieve 50\% net boil-off reduction for the perlite tank.
  This figure is the same as Figure~\ref{fig:power_A_II}, except that the $x$-axis has here been extended to included lower temperatures,
  and results for cooling configurations with DHC have been included.}
  \label{fig:power_A_II_eta_const}
\end{figure}

\begin{figure}
  \centering
  \includegraphics[width=0.45\textwidth]{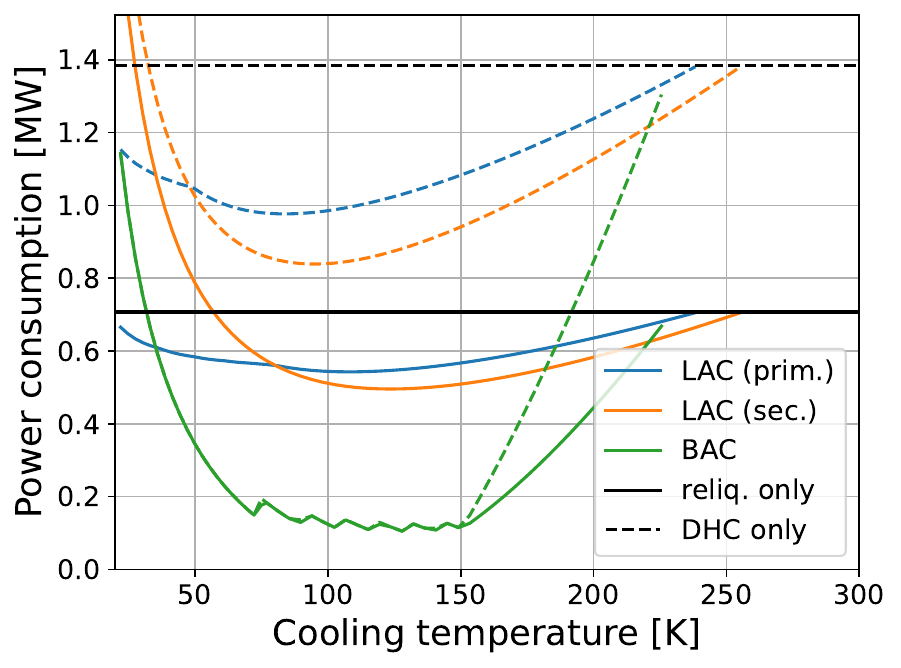}
  \caption{Power consumption required to achieve 50\% net boil-off reduction for the HePUR tank.
  This figure is the same as Figure~\ref{fig:power_B_II}, except that the $x$-axis has here been extended to included lower temperatures,
  and results for cooling configurations with DHC have been included.}
  \label{fig:power_B_II_eta_const}
\end{figure}

Figures~\ref{fig:power_A_I_eta_const}--\ref{fig:power_B_II_eta_const} are in direct correspondence to Figures~\ref{fig:power_A_I},~\ref{fig:power_A_II},~\ref{fig:power_B_I} and~\ref{fig:power_B_II}.
The solid lines correspond to cooling configurations 4--6 (reliquefaction only, LAC + reliquefaction, and BAC + reliquefaction),
which we considered also in the main text,
while the dashed lines correspond to analogous configuration with reliquefaction replaced by DHC.
It is immediately clear that, under the present assumptions on system efficiencies,
reliquefaction is significantly more power efficient than DHC. This can be explained as follows:

First, from Equation~\eqref{eq:power_reliq}, we observe that the reliquefaction power $P_{\mathrm{reliq}}$ is directly proportional to $\chi$, the energy required to reliquefy one kilogram of hydrogen.
Moreover, from Equation~\eqref{eq:power_dhc}, we observe that the DHC power $P_{\textsc{dhc}}$ is inversely proportional to the Carnot efficiency $\eta_{\textsc{dhc}}$.
Both $\chi$ and $\eta_{\textsc{dhc}}$ are associated with significant uncertainty, and future technological developments may alter their respective values.
Consequently, the sensitivity of the results presented herein to changes in $\chi$ and $\eta_{\textsc{dhc}}$ should be addressed.
To this end, observe that $P_{\textsc{dhc}}$ can be rewritten as
\begin{align}
  P_{\textsc{dhc}} &= \frac{\Delta H_{\mathrm{vap}} \mathrm{max}(\dot{m}_{\textsc{bog}} - \dot{m}_{\mathrm{fuel}}, 0)}{\eta_{\textsc{dhc}}\frac{T_{\mathrm{cool},\textsc{dhc}}}{T_{\mathrm{amb}} - T_{\mathrm{cool},\textsc{dhc}}}} \nonumber \\
  &=: \chi_{\textsc{dhc}} \mathrm{max}(\dot{m}_{\textsc{bog}} - \dot{m}_{\mathrm{fuel}}, 0) \nonumber \\
  &= \frac{\chi_{\textsc{dhc}}}{\chi} P_{\mathrm{reliq}}. \nonumber 
\end{align}
Hence, the power consumption of DHC
is the same as one would obtain for a reliquefaction unit
whose specific energy consumption is
\begin{align}
  \chi' &= \chi_{\textsc{dhc}} = \frac{\Delta H_{\mathrm{vap}}}{\eta_{\textsc{dhc}}\frac{T_{\mathrm{cool},\textsc{dhc}}}{T_{\mathrm{amb}} - T_{\mathrm{cool},\textsc{dhc}}}} \nonumber \\
  &= \frac{\SI{0.45}{\mega\joule\per\kilogram}}{0.2\frac{\SI{18}{\kelvin}}{\SI{282}{\kelvin}}} \approx \SI{9.7}{\kWh\per\kilogram}.\nonumber 
\end{align}
This is roughly double the specific energy consumption $\chi = \SI{5}{\kWh\per\kilogram}$ from the study by \citet{kim2023hrp} on hydrogen reliquefaction units.
It is then no surprise that reliquefaction compares favorably to DHC in Figures~\ref{fig:power_A_I_eta_const}--\ref{fig:power_B_II_eta_const}.

Given the above analogy between reliquefaction and DHC in terms of specific energy consumption,
interpolation between the results presented herein can be used to roughly estimate
the impact of reliquefaction units with different specific energy consumptions.
Conversely, it is also possible to define an effective Carnot efficiency for the reliquefaction unit:
\begin{align}
  \eta' &= \frac{\Delta H_{\mathrm{vap}}}{\chi\frac{T_{\mathrm{cool},\textsc{dhc}}}{T_{\mathrm{amb}} - T_{\mathrm{cool},\textsc{dhc}}}} \\
  &= \frac{\SI{0.45}{\mega\joule\per\kilogram}}{\SI{5}{\kWh\per\kilogram}\frac{\SI{18}{\kelvin}}{\SI{300}{\kelvin} - \SI{18}{\kelvin}}} \approx 0.39.\nonumber
\end{align}
With this analogy, one can similarly estimate the potential impact of increasing the refrigeration efficiency in DHC systems.

Another observation to be made from Figures~\ref{fig:power_A_I_eta_const}--\ref{fig:power_B_II_eta_const}
is that a minimum power consumption exists for all cooling setups, as claimed in the main text.
For certain setups, these minima occur below \SI{77}{\kelvin}, making them not visible in the figures of Section~\ref{sec:min_power}.
Nonetheless, it appears that LAC and BAC can reduce total power consumption even at $T_{\mathrm{cool}}$ close to the temperature of \lhto.
This indicates that the hitherto assumed Carnot efficiency of the LAC\slash BAC refrigeration units ($\eta = 0.5$, independent of $T_{\mathrm{cool}}$) is likely unreasonably high in the lowest temperature range.
Degradation of the Carnot efficiency at lower temperatures is considered in \ref{app:efficiency_degradation}.

\FloatBarrier

\section{Sensitivity to Degradation of Refrigeration Efficiency}
\label{app:efficiency_degradation}

Throughout this work, we have assumed that the Carnot efficiencies $\eta_{\textsc{lac}}$ and $\eta_{\textsc{bac}}$ of the refrigeration units
associated with LAC and BAC are independent of $T_{\mathrm{cool}}$.
However, in practice, one observes that such Carnot efficiencies tend to decrease for lower cooling temperatures.
In this section, we explore how degradation of the Carnot efficiency at lower cooling temperatures will impact power consumption.
To this end, instead of assuming $\eta_{\textsc{lac}} = \eta_{\textsc{bac}} = 0.5$ as before,
we now assume $\eta_{\textsc{lac}} = \eta_{\textsc{bac}} = \eta_T(T)$,
where $\eta_T$ is a piecewise linear interpolation between the following data points:
(\SI{22}{\kelvin}, 0.2)~\cite{deserranno2014ooa}, (\SI{77}{\kelvin}, 0.3)~\cite{walker2014cp1}, (\SI{150}{\kelvin}, 0.5)~\cite{chakravarthy2011aro}, and (\SI{285}{\kelvin}, 0.5)~\cite{zuhlsdorf2023hhp}.

\begin{figure}
  \centering
  \includegraphics[width=0.45\textwidth]{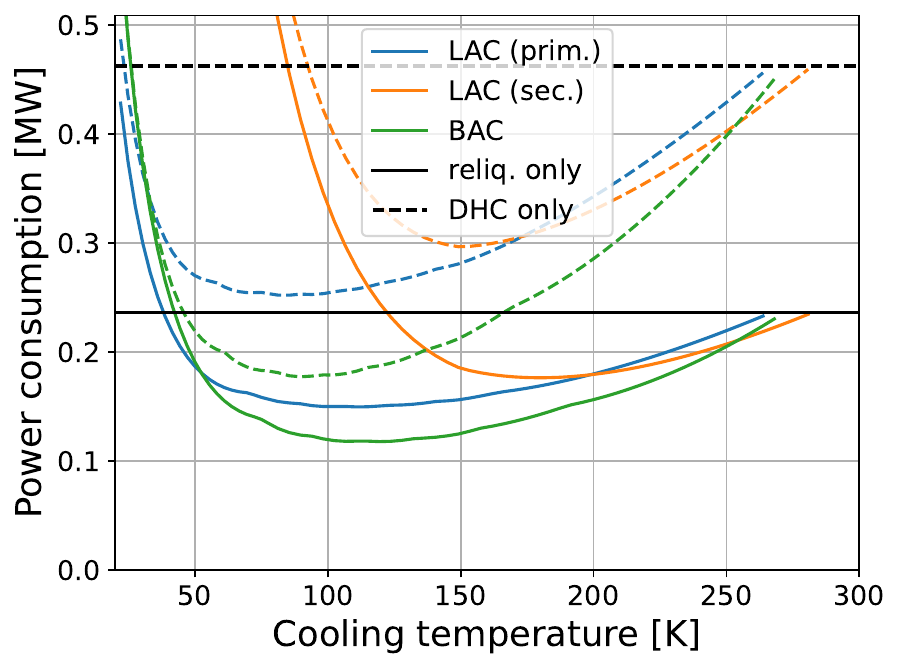}
  \caption{Power consumption required to achieve net zero boil-off for the perlite tank
  when degradation of refrigeration efficiency at low cooling temperatures is accounted for.
  Aside from the updated definitions of $\eta_{\textsc{lac}}$ and $\eta_{\textsc{bac}}$,
  this figure is the same as Figure~\ref{fig:power_A_I_eta_const}.}
  \label{fig:power_A_I_eta_interp}
\end{figure}

\begin{figure}
  \centering
  \includegraphics[width=0.45\textwidth]{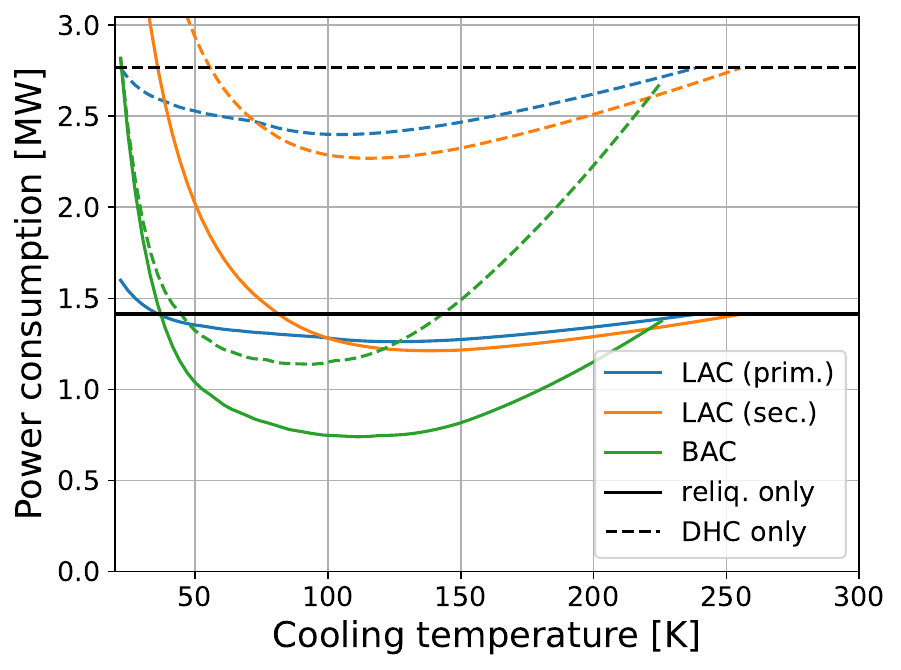}
  \caption{Power consumption required to achieve net zero boil-off for the HePUR tank
  when degradation of refrigeration efficiency at low cooling temperatures is accounted for.
  Aside from the updated definitions of $\eta_{\textsc{lac}}$ and $\eta_{\textsc{bac}}$,
  this figure is the same as Figure~\ref{fig:power_B_I_eta_const}.}
  \label{fig:power_B_I_eta_interp}
\end{figure}

Comparing Figures~\ref{fig:power_A_I_eta_interp} and~\ref{fig:power_B_I_eta_interp}
with Figures~\ref{fig:power_A_I_eta_const} and~\ref{fig:power_B_I_eta_const},
we observe that the Carnot efficiency degradation yields
a significant increase in power consumption at low $T_{\mathrm{cool}}$.
Consequently, pure reliquefaction becomes the most power efficient option
close to \lhto temperatures, as expected.
Additionally, for all configurations with LAC or BAC, 
the optimal $T_{\mathrm{cool}}$ become noticeable higher,
and cooling below \SI{100}{\kelvin} generally becomes unfavorable.
This shift also has some impact on the maximum achievable power savings,
especially for the BAC configurations which previously performed best at
$T_{\mathrm{cool}} \approx \SI{80}{\kelvin}$.
However, the authors believe this impact is minor enough that it does
not significantly distort the main findings of Section~\ref{sec:min_power}.
If anything, the main consequence of 
assuming constant Carnot efficiency is
a moderate underestimation of LAC's competetiveness in comparison to BAC.

\end{document}